\documentclass[useAMS,usenatbib]{mn2e}
\usepackage{graphicx}
\title[Ne~{\sc v} emission in five blue compact dwarf 
galaxies]
{The detection of $[$Ne~{\sc v}$]$ emission in five blue compact dwarf 
galaxies}
\author[Y. I. Izotov et al.]{Y.~I.~Izotov$^{1}$,
T.~X.~Thuan$^{2}$ and G.~Privon$^{2}$
\thanks{Based on observations
with the Multiple Mirror telescope (MMT). The MMT is operated by 
the MMT Observatory (MMTO), a joint venture of the Smithsonian Institution  
and the University of Arizona.}\\
$^{1}$Main Astronomical Observatory,
                     Ukrainian National Academy of Sciences,
                     Zabolotnoho 27, Kyiv 03680,  Ukraine\\
$^{2}$Astronomy Department, University of Virginia, P.O. 
                     Box 400325, Charlottesville, VA 22904, USA}
\begin{document}

\date{Accepted 1988 December 15. Received 1988 December 14; in original form 1988 October 11}

\pagerange{\pageref{firstpage}--\pageref{lastpage}} \pubyear{2012}

\maketitle

\label{firstpage}

\begin{abstract}
We report the discovery of the high-ionisation [Ne~{\sc v}] $\lambda$3426\AA\ emission 
line in the spectra of five blue compact dwarf
(BCD) galaxies. Adding the three previously known BCDs with 
[Ne~{\sc v}] emission, the entire sample of such galaxies now contains eight
objects. The detection of this line implies the presence of 
intense hard ionising radiation. Such radiation cannot be reproduced by models 
of high-mass X-ray binaries or massive stellar populations. 
Other mechanisms, such as AGN and/or fast radiative shocks, are needed.
We consider that fast radiative shocks is the most likely mechanism.
The observed 
[Ne~{\sc v}] $\lambda$3426/He~{\sc ii} $\lambda$4686 flux ratios in all 
eight galaxies can be reproduced by radiative shock models 
with shock velocities in the $\sim$ 300 -- 500 km s$^{-1}$ range, 
and with the shock ionising contribution being $\sim$10\% of the 
stellar ionising contribution. However, we cannot rule out that 
this 10\% part is produced by an AGN rather than by radiative shocks.  
\end{abstract}


\section[]{Introduction}\label{sec:intro}

Blue compact dwarf (BCD) galaxies are actively star-forming dwarf
galaxies in the local universe. They have a heavy element mass fraction $Z$ in
the range 1/30 -- 1/2 $Z_\odot$ \citep[e.g., ][]{IT99}, assuming a solar
oxygen abundance 12 + log O/H = 8.69 \citep{A09}. Thus, BCD massive
stellar populations have properties intermediate between
those of massive stars in solar-metallicity galaxies
and those of the first stars. The hardness of the ionising radiation in BCDs 
has long been
known to increase with decreasing metallicity \citep[e.g., ][]{C86}. BCDs  
constitute then excellent
nearby laboratories for studying high-ionisation emission
in a very metal-deficient environment.

The presence of hard radiation in BCDs 
is supported by the fact that strong nebular
He~{\sc ii} $\lambda$4686 emission is often seen in their spectra,
with a flux that increases with decreasing metallicity of the
ionised gas \citep{GIT00,TI05}. 
Besides He~{\sc ii} emission, high ionisation
emission lines of heavy elements ions are also seen in
the spectra of some BCDs. \citet{F01} and \citet{I01} first detected the 
high-ionisation [Fe~{\sc v}] $\lambda$4227 emission line in the BCDs  
Tol 1214$-$277 and SBS 0335$-$052E, respectively. The
presence of this line, just as that of the He~{\sc ii} $\lambda$4686 line, 
requires ionising radiation with photon energies in excess of 4 ryd (54.4 eV)
if ionised by radiation.
\citet{I01} discovered [Fe~{\sc vi}] -- [Fe~{\sc vii}]
emission in SBS 0335$-$052E, implying that this BCD contains
intense hard radiation, given that the ionisation potential
of Fe$^{5+}$ is 5.5 ryd and that of Fe$^{6+}$ is 7.3 ryd. Later, \citet{I04a}
and \citet{TI05} discovered [Ne~{\sc v}] $\lambda$3426 emission 
in the three BCDs Tol 1214$-$277, SBS 0335$-$052E and HS 0837$+$4717.
This line is often seen in Seyfert 2 galaxies, powered by the hard
non-thermal radiation of an active galactic nucleus (AGN). 
Furthermore, \citet{G98} and
\citet{A07} have also   
detected the mid-infrared (MIR) [Ne {\sc v}] $\lambda$14.3 $\mu$m and
$\lambda$24.3 $\mu$m emission lines in several massive galaxies, 
and have attributed them also to the presence of an AGN. 
On other hand, no MIR [Ne {\sc v}]
emission line has ever been detected in a BCD. In particular, all the three 
above BCDs with a detected [Ne {\sc v}] $\lambda$3426
emission line were observed with {\sl ISO} and {\sl Spitzer} 
\citep{T99,H04,W06,H10}, but no MIR [Ne {\sc v}] emission line
was seen, probably because of
the faintness of the lines and the insufficient sensitivity of the {\sl ISO} 
and {\sl Spitzer} space observatories.
The existence of the [Ne {\sc v}] $\lambda$3426 emission line requires
the presence of hard radiation with photon energies
above 7.1 ryd, i.e. in the extreme UV and soft X-ray range.
 Such hard ionising radiation is confirmed by the
detection of [Ne {\sc iv}] (the ionisation potential of Ne$^{3+}$ is 4.7 ryd)
and [Fe~{\sc vi}] -- [Fe~{\sc vii}] emission in  
Tol 1214$-$277 \citep{I04b}.

While the presence of hard radiation is well established in
some BCDs, the origin of this radiation is much less clear, in
spite of several attempts to account for it \citep[e.g., ][]{G91,SV98}.
Several mechanisms for producing hard ionising radiation have
been proposed, such as AGN \citep{IT08}, Wolf-Rayet stars \citep{S96}, 
high-mass X-ray binaries \citep{G91} and fast radiative shocks \citep{DS96}. 
However, no mechanism has emerged clearly as the leading candidate, 
in particular because of the lack of a large
database to confront models with observations. Despite the
importance of understanding the high-ionisation phenomenon
to interpret the spectra of primordial star-forming galaxies when
these are discovered in the future, very few observations of high ionisation
emission lines in metal-deficient BCDs exist. The He~{\sc ii}
$\lambda$4686 emission line has been detected in several hundred
BCDs \citep{GIT00,TI05,SB12}, while [Ne~{\sc v}] emission has been seen in
only three BCDs \citep{I04a,TI05}. The scarcity of data concerning 
[Ne~{\sc v}] emission 
is partly due to
the faintness of this high-ionisation line 
($<$3\% of H$\beta$).  
Its detection requires large telescope equipped with spectrographs 
operating efficiently in the near-UV range.  
To increase the number of known
BCDs with [Ne~{\sc v}] $\lambda$3426 emission, we
have embarked in a program to obtain high signal-to-noise spectra
in the blue wavelength region of a sample of BCDs with
the 6.5m MMT. These BCDs were selected from the Data Release 7 (DR7) of the 
Sloan
Digital Sky survey (SDSS) \citep{Ab09} to have  
spectra showing relatively strong He~{\sc ii} $\lambda$4686 emission. 
We describe the observational data in Section \ref{sec:obs}, and how  
element abundances are derived in Section \ref{sec:abund}. We
discuss possible mechanisms for the hard radiation in Section \ref{sec:hard}. 
Our conclusions are summarised in Section \ref{sec:concl}.

\begin{table}
 \centering
 \begin{minipage}{170mm}
  \caption{General characteristics of galaxies.}\label{tab1}
  \begin{tabular}{@{}lcccl@{}}
  \hline
Name        & R.A.      & Dec.     & $g$ &$M_g$ \\
            & J2000.0   & J2000.0  & mag & mag      \\
 \hline
J0905$+$0335&09:05:31.09    &$+$03:35:30.38&17.6&$-$18.4           \\ 
J0920$+$5234&09:20:56.08    &$+$52:34:04.32&15.7&$-$16.8 \\ 
J1016$+$3754&10:16:24.53    &$+$37:54:45.97&15.9&$-$15.1 \\ 
J1044$+$0353&10:44:57.80    &$+$03:53:13.15&17.5&$-$16.0                \\ 
J1050$+$1538&10:50:32.51    &$+$15:38:06.31&18.2&$-$19.4                \\ 
J1053$+$5016&10:53:10.82    &$+$50:16:53.21&19.9&$-$11.4                \\ 
J1230$+$1202&12:30:48.60    &$+$12:02:42.82&16.7&$-$14.4 \\ 
J1323$-$0132&13:23:47.47    &$-$01:32:51.95&18.1&$-$16.7                \\ 
J1423$+$2257&14:23:42.88    &$+$22:57:28.79&17.9&$-$17.7                \\ 
J1426$+$3822&14:26:28.18    &$+$38:22:58.67&18.2&$-$16.6                \\ 
J1448$-$0110&14:48:05.36    &$-$01:10:57.71&16.4&$-$18.8                \\ 
J1545$+$0858&15:45:43.55    &$+$08:58:01.35&16.9&$-$19.0                \\
\hline
\end{tabular}
\end{minipage}
\end{table}

\begin{figure}
\hbox{
\includegraphics[width=4.0cm,angle=0.]{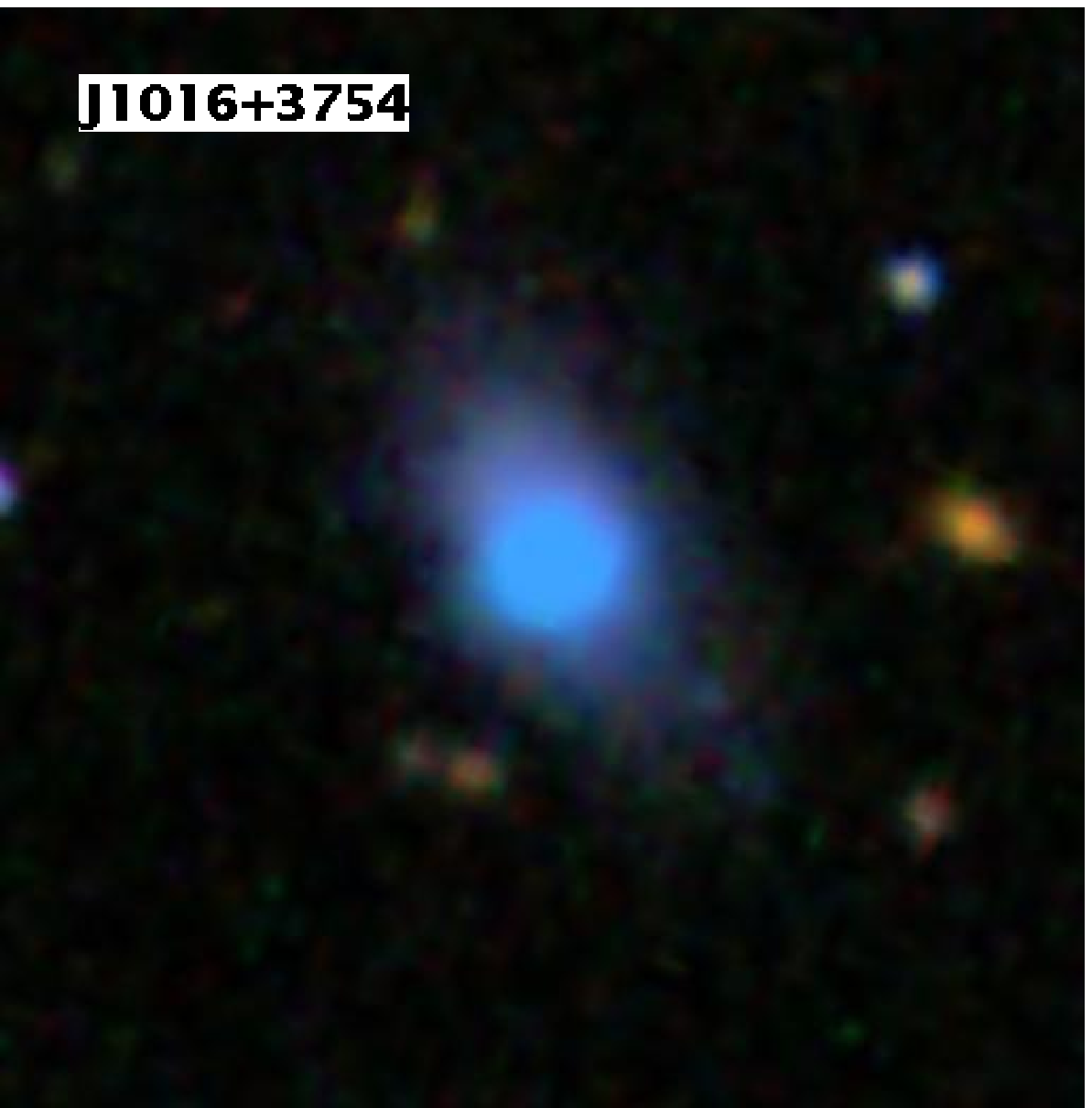}
\hspace{0.1cm}\includegraphics[width=4.0cm,angle=0.]{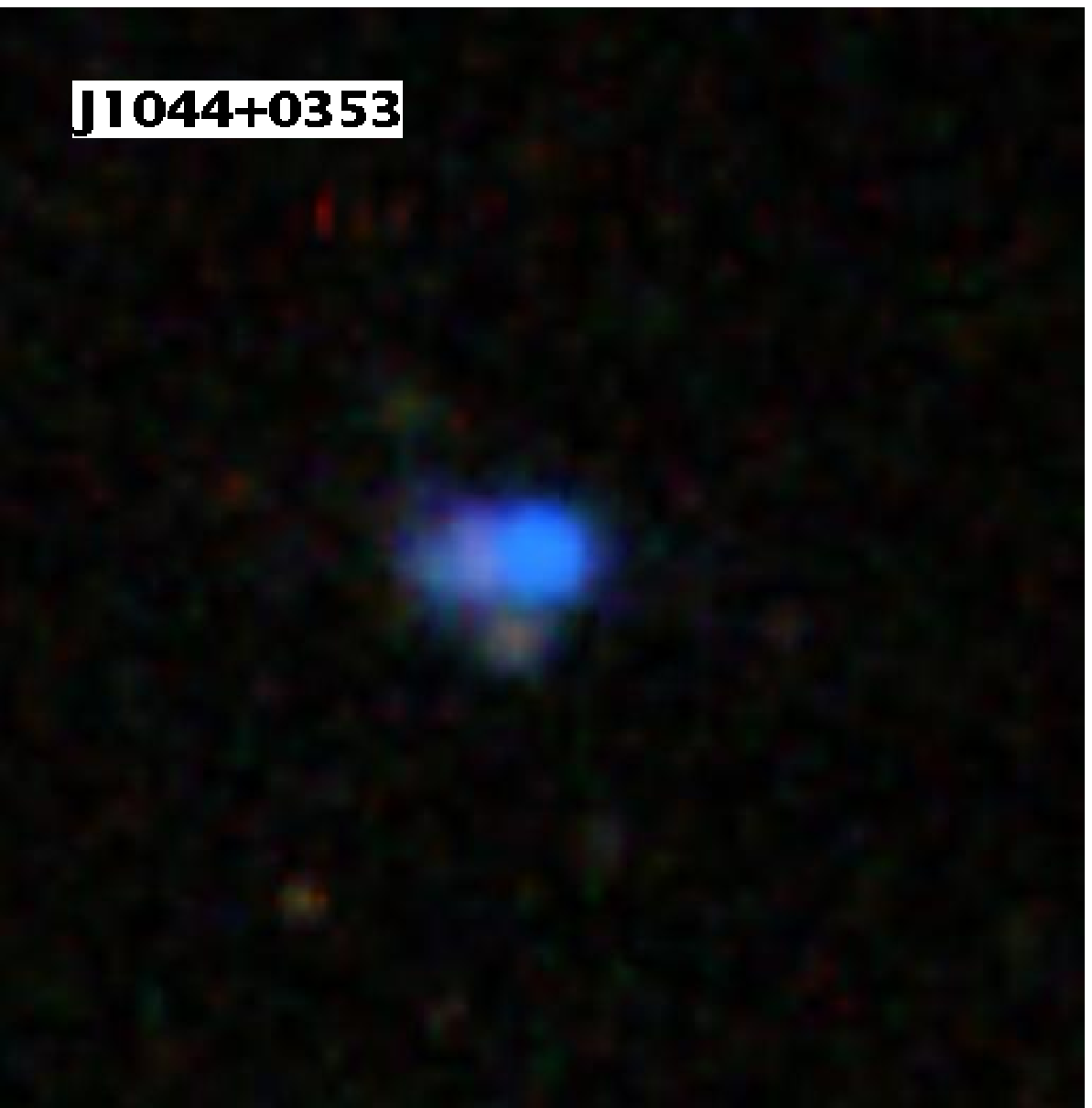}
}
\vspace{0.15cm}
\hbox{
\includegraphics[width=4.0cm,angle=0.]{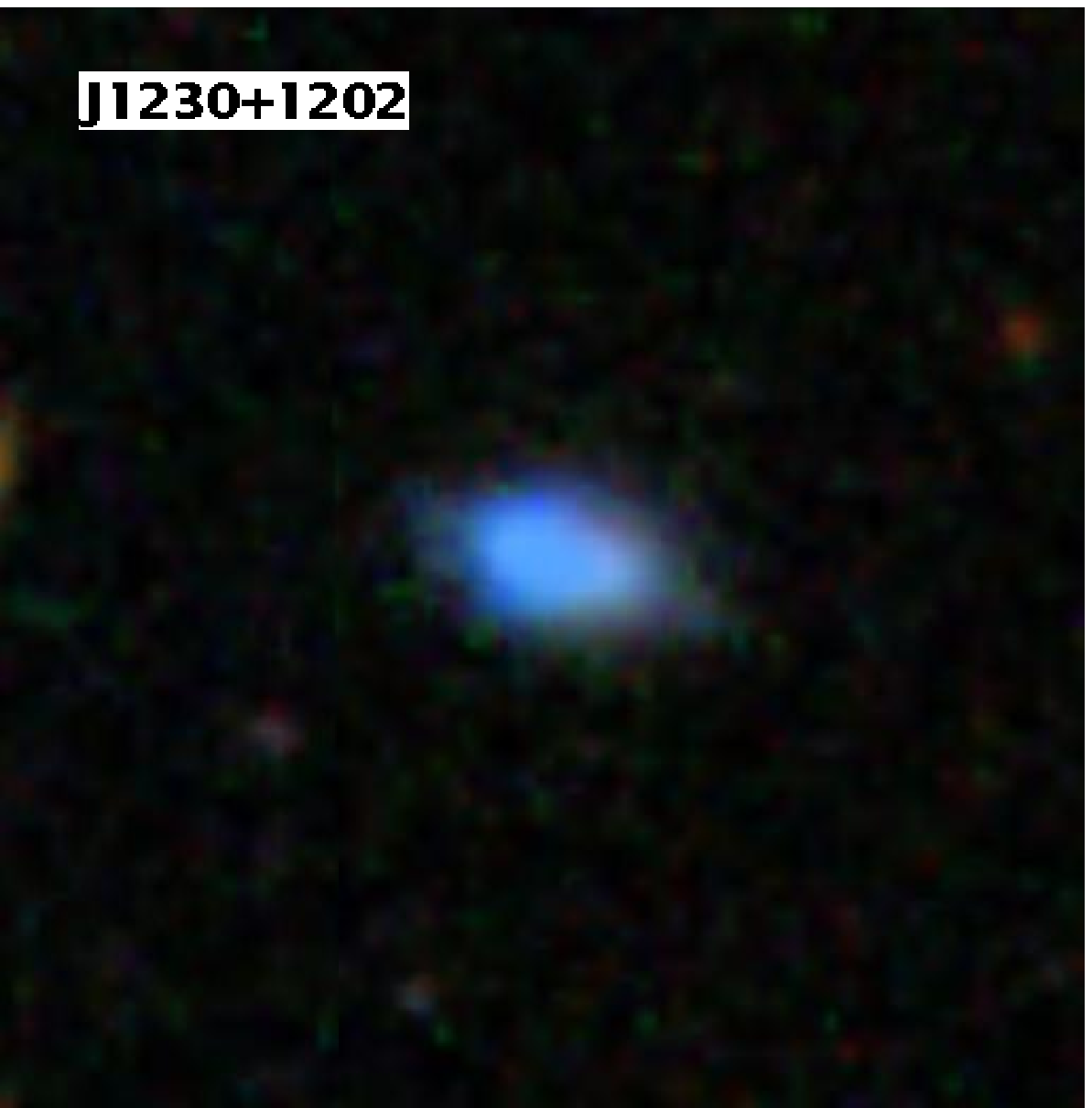}
\hspace{0.1cm}\includegraphics[width=4.0cm,angle=0.]{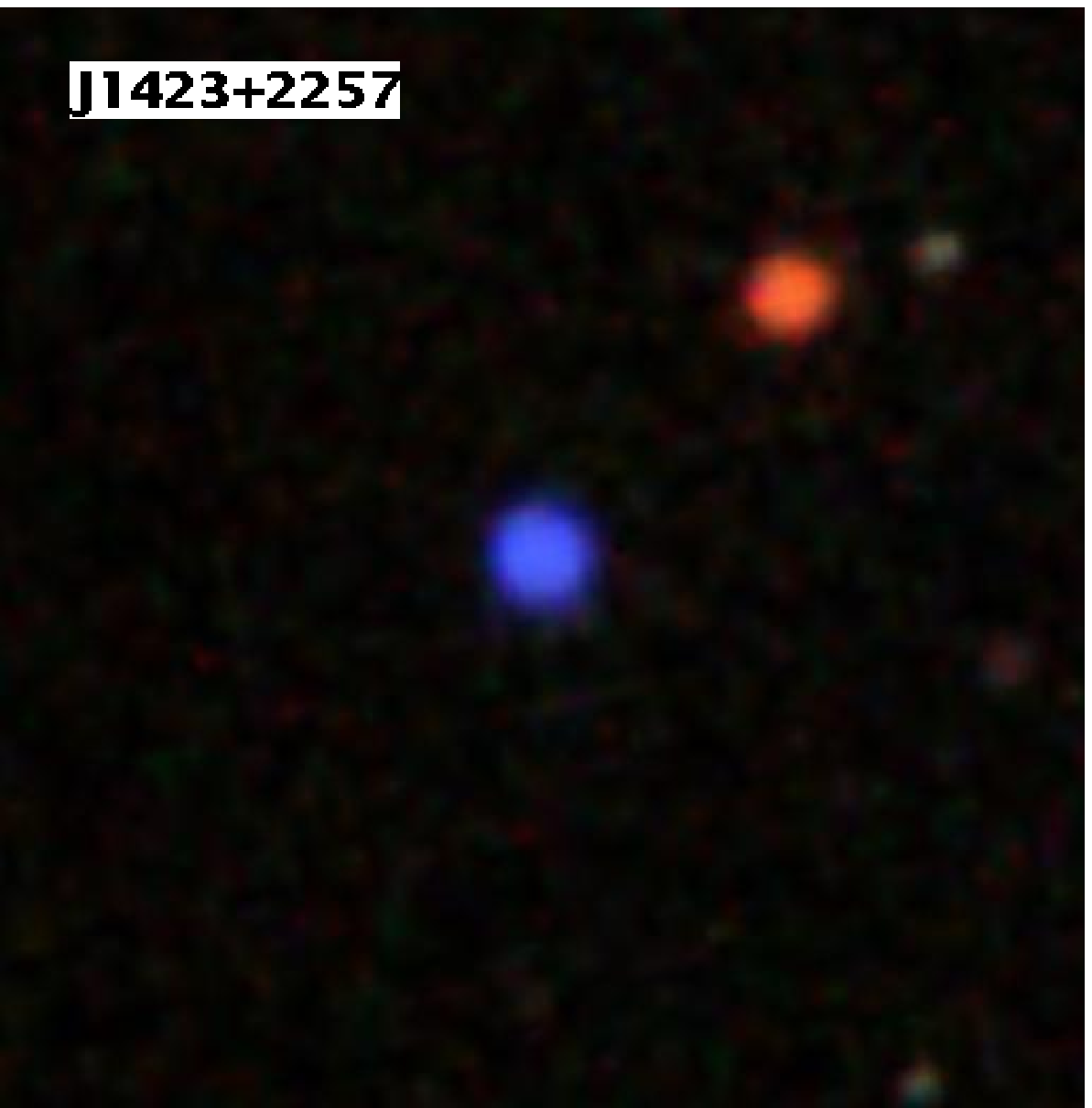}
}
\vspace{0.15cm}
\hbox{
\includegraphics[width=4.0cm,angle=0.]{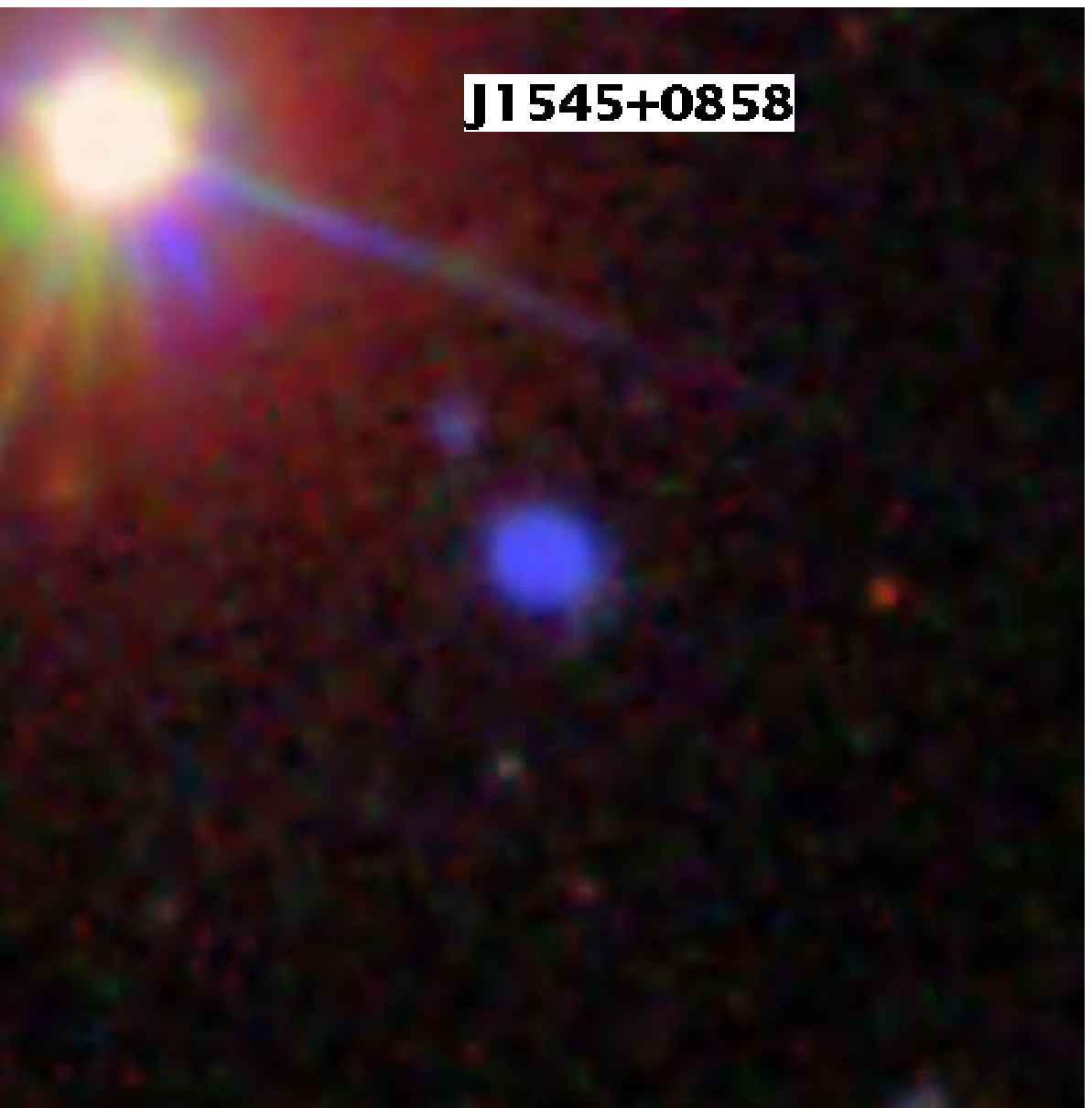}
}
 \caption{50\arcsec$\times$50\arcsec\ SDSS images of galaxies with 
[Ne~{\sc v}] $\lambda$3426\AA\
emission.}
\label{fig1}
\end{figure}

\begin{figure*}
\includegraphics[width=12.cm,angle=-90.]{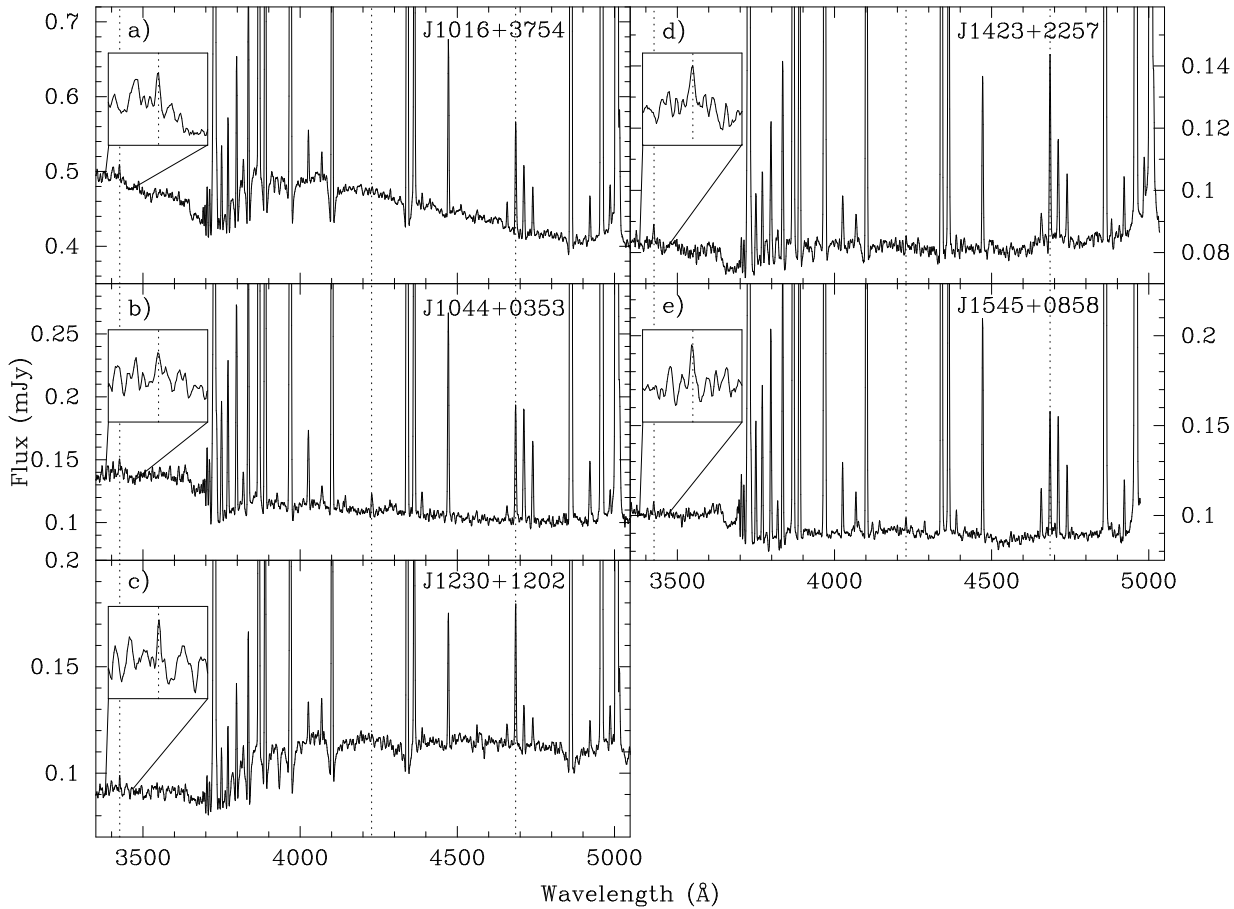}
 \caption{Redshift-corrected spectra of galaxies with  
[Ne~{\sc v}] $\lambda$3426\AA\ emission. The vertical dotted lines show 
the location (from left to right) of the [Ne~{\sc v}] $\lambda$3426\AA,
[Fe~{\sc v}] $\lambda$4227\AA, and He~{\sc ii} $\lambda$4686\AA\ emission
lines. The insets show expanded parts of the spectra in the wavelength
range $\lambda$$\lambda$3380--3470\AA. The  
[Ne~{\sc v}] $\lambda$3426\AA\ emission line is marked by a vertical dotted 
line.}
\label{fig2}
\end{figure*}

\begin{table*}
 \centering
 \begin{minipage}{170mm}
  \caption{Extinction-corrected emission line fluxes.$^{\rm a}$}\label{tab2}
  \begin{tabular}{@{}lrrrrrr@{}} \hline
Line                         &J0905+0335      &J0920+5234      &J1016+3754      &J1044+0353      &J1050+1538      &J1053+5016      \\
  \hline
3188 He {\sc i}              &  3.24$\pm$0.71 &  4.43$\pm$0.71 &  1.82$\pm$0.39 &  3.44$\pm$0.30 &  ...~~~~~      &  2.07$\pm$0.46 \\
3203 He {\sc ii}             &  1.78$\pm$0.57 &  ...~~~~~      &  ...~~~~~      &  0.82$\pm$0.17 &  ...~~~~~      &  ...~~~~~      \\
3426 [Ne {\sc v}]$^{\rm b}$  &  $<$0.55~~     &  $<$0.25~~     &  0.58$\pm$0.15 &  0.61$\pm$0.15 &  $<$0.23~~     &  $<$0.17~~     \\
3712 H15                     &  1.42$\pm$0.38 &  2.91$\pm$0.41 &  1.47$\pm$0.16 &  1.77$\pm$0.18 &  1.84$\pm$0.26 &  2.61$\pm$0.31 \\
3727 [O {\sc ii}]            &104.24$\pm$1.73 &128.98$\pm$2.02 & 63.96$\pm$0.98 & 30.86$\pm$0.51 &119.78$\pm$1.88 &145.59$\pm$2.24 \\
3750 H12                     &  3.06$\pm$0.38 &  3.83$\pm$0.35 &  2.25$\pm$0.16 &  3.35$\pm$0.16 &  3.60$\pm$0.29 &  3.34$\pm$0.25 \\
3770 H11                     &  3.47$\pm$0.39 &  4.54$\pm$0.32 &  3.03$\pm$0.17 &  4.03$\pm$0.16 &  4.07$\pm$0.29 &  4.28$\pm$0.25 \\
3797 H10                     &  5.04$\pm$0.40 &  5.56$\pm$0.32 &  4.16$\pm$0.18 &  5.63$\pm$0.17 &  5.78$\pm$0.29 &  5.29$\pm$0.24 \\
3820 He {\sc i}              &  0.66$\pm$0.23 &  1.03$\pm$0.13 &  0.73$\pm$0.10 &  0.77$\pm$0.06 &  1.08$\pm$0.16 &  0.59$\pm$0.11 \\
3835 H9                      &  6.87$\pm$0.39 &  7.87$\pm$0.30 &  6.38$\pm$0.18 &  7.61$\pm$0.19 &  7.86$\pm$0.29 &  7.43$\pm$0.25 \\
3868 [Ne {\sc iii}]          & 46.62$\pm$0.82 & 42.84$\pm$0.69 & 36.94$\pm$0.57 & 35.68$\pm$0.57 & 55.55$\pm$0.90 & 48.52$\pm$0.76 \\
3889 He {\sc i}+H8           & 21.12$\pm$0.52 & 19.87$\pm$0.41 & 19.77$\pm$0.34 & 21.11$\pm$0.36 & 20.49$\pm$0.42 & 19.58$\pm$0.37 \\
3968 [Ne {\sc iii}]+H7       & 30.99$\pm$0.66 & 28.01$\pm$0.51 & 26.83$\pm$0.43 & 27.46$\pm$0.45 & 33.85$\pm$0.60 & 30.47$\pm$0.52 \\
4026 He {\sc i}              &  1.32$\pm$0.20 &  1.73$\pm$0.14 &  1.15$\pm$0.08 &  1.75$\pm$0.09 &  1.75$\pm$0.15 &  1.24$\pm$0.11 \\
4068 [S {\sc ii}]            &  1.13$\pm$0.24 &  1.13$\pm$0.11 &  0.62$\pm$0.08 &  0.73$\pm$0.08 &  1.52$\pm$0.17 &  1.52$\pm$0.12 \\
4076 [S {\sc ii}]            &  0.52$\pm$0.24 &  0.38$\pm$0.11 &  ...~~~~~      &  ...~~~~~      &  ...~~~~~      &  0.40$\pm$0.10 \\
4101 H$\delta$               & 27.26$\pm$0.56 & 26.02$\pm$0.47 & 26.28$\pm$0.41 & 26.39$\pm$0.42 & 27.13$\pm$0.49 & 26.16$\pm$0.44 \\
4227 [Fe {\sc v}]            &  ...~~~~~      &  ...~~~~~      &  ...~~~~~      &  0.47$\pm$0.06 &  ...~~~~~      &  ...~~~~~      \\
4340 H$\gamma$               & 49.32$\pm$0.85 & 47.25$\pm$0.74 & 48.18$\pm$0.71 & 47.61$\pm$0.72 & 46.91$\pm$0.75 & 47.18$\pm$0.72 \\
4363 [O {\sc iii}]           & 11.84$\pm$0.34 &  9.70$\pm$0.20 & 11.88$\pm$0.20 & 13.97$\pm$0.23 & 12.23$\pm$0.27 &  9.96$\pm$0.20 \\
4388 He {\sc i}              &  ...~~~~~      &  ...~~~~~      &  0.36$\pm$0.08 &  0.63$\pm$0.10 &  0.53$\pm$0.11 &  0.36$\pm$0.08 \\
4471 He {\sc i}              &  3.89$\pm$0.25 &  3.70$\pm$0.15 &  3.22$\pm$0.10 &  3.80$\pm$0.10 &  3.50$\pm$0.14 &  3.66$\pm$0.12 \\
4658 [Fe {\sc iii}]          &  0.77$\pm$0.19 &  0.85$\pm$0.10 &  0.49$\pm$0.07 &  0.40$\pm$0.06 &  0.95$\pm$0.11 &  0.37$\pm$0.06 \\
4686 He {\sc ii}             &  1.92$\pm$0.22 &  0.97$\pm$0.10 &  2.10$\pm$0.09 &  2.01$\pm$0.08 &  1.59$\pm$0.14 &  1.78$\pm$0.10 \\
4711 [Ar {\sc iv}]+He {\sc i}&  1.63$\pm$0.17 &  1.22$\pm$0.12 &  1.49$\pm$0.09 &  2.16$\pm$0.09 &  1.36$\pm$0.13 &  1.17$\pm$0.09 \\
4740 [Ar {\sc iv}]           &  1.46$\pm$0.19 &  0.63$\pm$0.10 &  0.87$\pm$0.08 &  1.34$\pm$0.06 &  0.91$\pm$0.12 &  0.68$\pm$0.08 \\
4861 H$\beta $               &100.00$\pm$1.55 &100.00$\pm$1.47 &100.00$\pm$1.43 &100.00$\pm$1.45 &100.00$\pm$1.47 &100.00$\pm$1.46 \\
4921 He {\sc i}              &  0.97$\pm$0.19 &  1.22$\pm$0.12 &  0.78$\pm$0.06 &  1.10$\pm$0.06 &  1.30$\pm$0.13 &  0.93$\pm$0.08 \\
4959 [O {\sc iii}]           &175.43$\pm$2.66 &181.47$\pm$2.65 &146.09$\pm$2.09 &141.58$\pm$2.05 &216.01$\pm$3.14 &198.09$\pm$2.87 \\
5007 [O {\sc iii}]           &526.29$\pm$6.55 &541.11$\pm$7.86 &428.87$\pm$6.11 &415.16$\pm$5.98 &644.11$\pm$9.34 &586.88$\pm$8.47 \\
5015 He {\sc i}              &  ...~~~~~      &  ...~~~~~      &  2.24$\pm$0.08 &  ...~~~~~      &  ...~~~~~      &  2.07$\pm$0.11 \\
$C$(H$\beta$)                &0.000           &0.000           &0.000           &0.020           &0.180           &0.000           \\
$F$(H$\beta$)$^{\rm c}$      &75.87           &201.20          &474.50          &327.50          &182.87          &253.80          \\
EW(H$\beta$), \AA\           &116.4           &139.1           &96.7            &257.8           &198.5           &127.8           \\
EW(abs), \AA\                &0.0             &1.4             &0.0             &0.8             &0.0             &0.9             \\
\hline
\end{tabular}

$^{\rm a}$Emission-line fluxes are in 100$\times$$I$($\lambda$)/$I$(H$\beta$).

$^{\rm b}$A 1$\sigma$ upper limit is given when the [Ne~{\sc v}] $\lambda$3426 
emission line is not detected.

$^{\rm c}$Observed H$\beta$ flux in 10$^{-16}$ erg s$^{-1}$cm$^{-2}$.

\end{minipage}
\end{table*}

\setcounter{table}{1}

\begin{table*}
 \centering
 \begin{minipage}{170mm}
  \caption{---{\sl Continued.}$^{\rm a}$}
  \begin{tabular}{@{}lrrrrrr@{}} \hline
Line                         &J1230+1202      &J1323$-$0132    &J1423+2257      &J1426+3822      &J1448$-$0110    &J1545+0858      \\
  \hline
3188 He {\sc i}              &  2.75$\pm$1.11 &  ...~~~~~      &  4.15$\pm$0.59 &  4.15$\pm$0.77 &  2.83$\pm$0.56 &  3.33$\pm$0.22 \\
3203 He {\sc ii}             &  2.87$\pm$0.81 &  ...~~~~~      &  ...~~~~~      &  ...~~~~~      &  ...~~~~~      &  1.09$\pm$0.16 \\
3426 [Ne {\sc v}]$^{\rm b}$  &  0.79$\pm$0.36 &  $<$0.39~~     &  0.91$\pm$0.31 &  $<$0.31~~     &  $<$0.20~~     &  0.51$\pm$0.16 \\
3712 H15                     &  1.29$\pm$0.38 &  ...~~~~~      &  1.36$\pm$0.29 &  1.36$\pm$0.32 &  2.51$\pm$0.53 &  1.93$\pm$0.15 \\
3727 [O {\sc ii}]            &118.38$\pm$1.91 & 21.82$\pm$0.73 & 83.15$\pm$1.38 & 85.74$\pm$1.42 & 97.03$\pm$1.61 & 69.97$\pm$1.08 \\
3750 H12                     &  2.34$\pm$0.36 &  3.36$\pm$0.80 &  2.70$\pm$0.34 &  2.61$\pm$0.32 &  3.63$\pm$0.51 &  3.35$\pm$0.15 \\
3770 H11                     &  3.17$\pm$0.37 &  4.12$\pm$0.70 &  3.12$\pm$0.32 &  3.45$\pm$0.37 &  4.19$\pm$0.45 &  4.08$\pm$0.15 \\
3797 H10                     &  4.09$\pm$0.38 &  5.51$\pm$0.70 &  5.07$\pm$0.34 &  4.96$\pm$0.36 &  5.91$\pm$0.47 &  5.35$\pm$0.16 \\
3820 He {\sc i}              &  0.93$\pm$0.21 &  1.56$\pm$0.45 &  ...~~~~~      &  0.60$\pm$0.17 &  ...~~~~~      &  0.96$\pm$0.08 \\
3835 H9                      &  6.52$\pm$0.38 &  7.18$\pm$0.69 &  6.52$\pm$0.32 &  6.56$\pm$0.35 &  7.43$\pm$0.45 &  7.19$\pm$0.18 \\
3868 [Ne {\sc iii}]          & 41.53$\pm$0.73 & 55.35$\pm$1.13 & 42.47$\pm$0.73 & 47.70$\pm$0.81 & 53.28$\pm$0.91 & 41.56$\pm$0.65 \\
3889 He {\sc i}+H8           & 19.93$\pm$0.50 & 19.06$\pm$0.77 & 20.71$\pm$0.47 & 19.64$\pm$0.47 & 19.85$\pm$0.48 & 19.00$\pm$0.32 \\
3968 [Ne {\sc iii}]+H7       & 27.89$\pm$0.57 & 33.38$\pm$0.90 & 28.87$\pm$0.56 & 29.87$\pm$0.60 & 31.30$\pm$0.63 & 29.27$\pm$0.47 \\
4026 He {\sc i}              &  1.39$\pm$0.22 &  1.44$\pm$0.26 &  1.93$\pm$0.19 &  1.72$\pm$0.19 &  1.80$\pm$0.21 &  1.59$\pm$0.09 \\
4068 [S {\sc ii}]            &  1.34$\pm$0.19 &  ...~~~~~      &  1.18$\pm$0.18 &  0.80$\pm$0.12 &  0.99$\pm$0.17 &  0.85$\pm$0.06 \\
4076 [S {\sc ii}]            &  ...~~~~~      &  ...~~~~~      &  ...~~~~~      &  ...~~~~~      &  ...~~~~~      &  0.24$\pm$0.04 \\
4101 H$\delta$               & 26.36$\pm$0.55 & 26.08$\pm$0.73 & 27.42$\pm$0.51 & 25.50$\pm$0.52 & 26.01$\pm$0.53 & 26.33$\pm$0.42 \\
4227 [Fe {\sc v}]            &  ...~~~~~      &  ...~~~~~      &  0.47$\pm$0.14 &  ...~~~~~      &  ...~~~~~      &  0.33$\pm$0.05 \\
4340 H$\gamma$               & 49.50$\pm$0.82 & 47.79$\pm$0.94 & 46.89$\pm$0.75 & 48.75$\pm$0.81 & 47.06$\pm$0.78 & 47.35$\pm$0.71 \\
4363 [O {\sc iii}]           & 11.04$\pm$0.29 & 20.80$\pm$0.52 & 13.52$\pm$0.28 & 12.35$\pm$0.29 &  9.51$\pm$0.26 & 13.33$\pm$0.22 \\
4388 He {\sc i}              &  0.57$\pm$0.21 &  ...~~~~~      &  ...~~~~~      &  ...~~~~~      &  ...~~~~~      &  0.44$\pm$0.08 \\
4471 He {\sc i}              &  3.55$\pm$0.19 &  3.80$\pm$0.30 &  3.52$\pm$0.15 &  3.89$\pm$0.24 &  3.51$\pm$0.19 &  3.60$\pm$0.10 \\
4658 [Fe {\sc iii}]          &  0.80$\pm$0.16 &  ...~~~~~      &  0.73$\pm$0.12 &  0.89$\pm$0.16 &  1.00$\pm$0.15 &  0.87$\pm$0.07 \\
4686 He {\sc ii}             &  3.81$\pm$0.20 &  1.37$\pm$0.33 &  3.23$\pm$0.15 &  2.44$\pm$0.19 &  0.86$\pm$0.13 &  2.04$\pm$0.08 \\
4711 [Ar {\sc iv}]+He {\sc i}&  1.53$\pm$0.21 &  4.71$\pm$0.31 &  1.84$\pm$0.13 &  1.82$\pm$0.16 &  1.80$\pm$0.18 &  1.88$\pm$0.07 \\
4740 [Ar {\sc iv}]           &  0.92$\pm$0.24 &  3.98$\pm$0.39 &  1.21$\pm$0.14 &  1.25$\pm$0.18 &  0.95$\pm$0.14 &  0.97$\pm$0.06 \\
4861 H$\beta $               &100.00$\pm$1.51 &100.00$\pm$1.69 &100.00$\pm$1.49 &100.00$\pm$1.52 &100.00$\pm$1.52 &100.00$\pm$1.45 \\
4921 He {\sc i}              &  0.80$\pm$0.15 &  0.93$\pm$0.28 &  0.95$\pm$0.13 &  1.22$\pm$0.21 &  1.00$\pm$0.19 &  0.79$\pm$0.07 \\
4959 [O {\sc iii}]           &155.06$\pm$2.30 &225.18$\pm$3.59 &169.71$\pm$2.50 &189.25$\pm$2.82 &214.47$\pm$3.19 &177.26$\pm$2.56 \\
5007 [O {\sc iii}]           &455.52$\pm$6.67 &635.96$\pm$9.94 &514.05$\pm$7.52 &548.03$\pm$8.07 &625.41$\pm$9.24 &530.30$\pm$6.24 \\
5015 He {\sc i}              &  1.90$\pm$0.17 &  ...~~~~~      &  ...~~~~~      &  ...~~~~~      &  ...~~~~~      &  ...~~~~~      \\
$C$(H$\beta$)                &0.000           &0.000           &0.400           &0.000           &0.135           &0.110           \\
$F$(H$\beta$)$^{\rm c}$      &107.70          &63.32           &133.80          &80.16           &337.70          &235.80          \\
EW(H$\beta$), \AA\           &84.5            &264.7           &129.0           &106.8           &134.7           &211.8           \\
EW(abs), \AA\                &0.0             &0.4             &0.0             &0.0             &1.1             &0.6             \\
\hline
\end{tabular}

$^{\rm a}$Emission-line fluxes are in units 100$\times$$I$($\lambda$)/$I$(H$\beta$).

$^{\rm b}$A 1$\sigma$ upper limit is given when the [Ne~{\sc v}] $\lambda$3426 
emission line is not detected.

$^{\rm b}$Observed H$\beta$ flux in 10$^{-16}$ erg s$^{-1}$cm$^{-2}$.

\end{minipage}
\end{table*}

\begin{table*}
 \centering
 \begin{minipage}{170mm}
  \caption{Physical conditions and element abundances.}\label{tab3}
  \begin{tabular}{@{}lrrrrrr@{}} \hline
Property                          &J0905+0335      &J0920+5234      &J1016+3754       &J1044+0353     &J1050+1538       &J1053+5016      \\ 
\hline
  $T_{\rm e}$(O {\sc iii}) (K)    &16103$\pm$237   &14498$\pm$150   &17851$\pm$196   &19928$\pm$243   &14875$\pm$165   &14153$\pm$138    \\
  $T_{\rm e}$(O {\sc ii}) (K)     &14924$\pm$200   &13931$\pm$133   &15718$\pm$156   &16272$\pm$176   &14187$\pm$144   &13685$\pm$123    \\
\\
  O$^+$/H$^+$ ($\times$10$^4$)    &0.096$\pm$0.004 &0.148$\pm$0.005 &0.050$\pm$0.001 &0.022$\pm$0.001 &0.130$\pm$0.004 &0.177$\pm$0.053  \\
  O$^{2+}$/H$^+$ ($\times$10$^4$) &0.486$\pm$0.019 &0.650$\pm$0.019 &0.314$\pm$0.009 &0.240$\pm$0.007 &0.724$\pm$0.023 &0.752$\pm$0.021  \\
  O$^{3+}$/H$^+$ ($\times$10$^6$) &1.147$\pm$0.134 &0.847$\pm$0.091 &0.941$\pm$0.047 &0.542$\pm$0.027 &1.563$\pm$0.141 &1.838$\pm$0.112  \\
  O/H ($\times$10$^4$)            &0.594$\pm$0.019 &0.806$\pm$0.020 &0.374$\pm$0.009 &0.268$\pm$0.007 &0.869$\pm$0.023 &0.947$\pm$0.022  \\
  12 + log(O/H)                   &7.77$\pm$0.01   &7.91$\pm$0.01   &7.57$\pm$0.01   &7.43$\pm$0.01   &7.94$\pm$0.01   &7.98$\pm$0.01    \\
\\
  Ne$^{++}$/H$^+$ ($\times$10$^5$)&1.020$\pm$0.042 &1.267$\pm$0.041 &0.616$\pm$0.018 &0.456$\pm$0.014 &1.522$\pm$0.051 &1.544$\pm$0.048  \\
  ICF                             &1.0587          &1.0632          &1.0515          &1.0342          &1.0539          &1.0677           \\
  log(Ne/O)                       &$-$0.74$\pm$0.02&$-$0.78$\pm$0.02&$-$0.76$\pm$0.02&$-$0.75$\pm$0.02&$-$0.73$\pm$0.02&$-$0.76$\pm$0.02 \\
\\
  Fe$^{++}$/H$^+$ ($\times$10$^6$)&0.148$\pm$0.037 &0.194$\pm$0.023 &0.083$\pm$0.012 &0.063$\pm$0.010 &0.208$\pm$0.025 &0.089$\pm$0.015  \\
  ICF                             &8.4843          &7.4619          &10.2202         &16.8769         &9.2241          &7.3197           \\
  log(Fe/O)                       &$-$1.67$\pm$0.11&$-$1.74$\pm$0.05&$-$1.64$\pm$0.07&$-$1.40$\pm$0.07&$-$1.66$\pm$0.05&$-$2.16$\pm$0.07 \\
\hline
Property                          &J1230+1202      &J1323$-$0132    &J1423+2257       &J1426+3822     &J1448$-$0110    &J1545+0858      \\ 
\hline
  $T_{\rm e}$(O {\sc iii}) (K)    &16659$\pm$239   &19485$\pm$323   &17440$\pm$217   &16046$\pm$205   &13503$\pm$167   &17009$\pm$169    \\
  $T_{\rm e}$(O {\sc ii}) (K)     &15209$\pm$198   &16190$\pm$239   &15558$\pm$175   &14893$\pm$174   &13189$\pm$151   &15373$\pm$138    \\
\\
  O$^+$/H$^+$ ($\times$10$^4$)    &0.103$\pm$0.004 &0.016$\pm$0.001 &0.067$\pm$0.002 &0.079$\pm$0.003 &0.134$\pm$0.005 &0.058$\pm$0.002  \\
  O$^{2+}$/H$^+$ ($\times$10$^4$) &0.390$\pm$0.014 &0.390$\pm$0.015 &0.394$\pm$0.013 &0.515$\pm$0.018 &0.912$\pm$0.033 &0.431$\pm$0.011  \\
  O$^{3+}$/H$^+$ ($\times$10$^6$) &2.107$\pm$0.126 &0.573$\pm$0.141 &1.680$\pm$0.091 &1.493$\pm$0.127 &1.047$\pm$0.166 &1.102$\pm$0.049  \\
  O/H ($\times$10$^4$)            &0.514$\pm$0.015 &0.411$\pm$0.016 &0.478$\pm$0.013 &0.609$\pm$0.018 &1.056$\pm$0.033 &0.501$\pm$0.011  \\
  12 + log(O/H)                   &7.71$\pm$0.01   &7.61$\pm$0.02   &7.68$\pm$0.01   &7.78$\pm$0.01   &8.02$\pm$0.01   &7.70$\pm$0.01    \\
\\
  Ne$^{2+}$/H$^+$ ($\times$10$^5$)&0.829$\pm$0.032 &0.745$\pm$0.030 &0.751$\pm$0.025 &1.054$\pm$0.038 &1.963$\pm$0.078 &0.785$\pm$0.022  \\
  ICF                             &1.0809          &1.0215          &1.0570          &1.0499          &1.0463          &1.0451           \\
  log(Ne/O)                       &$-$0.76$\pm$0.02&$-$0.73$\pm$0.02&$-$0.78$\pm$0.02&$-$0.74$\pm$0.02&$-$0.71$\pm$0.02&$-$0.79$\pm$0.02 \\
\\
  Fe$^{2+}$/H$^+$ ($\times$10$^6$)&0.148$\pm$0.029 & ...~~~~~~      &0.127$\pm$0.021 &0.173$\pm$0.031 &0.265$\pm$0.040 &0.156$\pm$0.012  \\
  ICF                             &6.8370          & ...~~~~~~      &9.7400          &10.5541         &10.8730         &11.7320          \\
  log(Fe/O)                       &$-$1.71$\pm$0.09& ...~~~~~~      &$-$1.59$\pm$0.08&$-$1.52$\pm$0.08&$-$1.56$\pm$0.07&$-$1.44$\pm$0.03 \\
\hline
\end{tabular}
\end{minipage}
\end{table*}

\section[]{MMT observations and data reduction}\label{sec:obs}

We have obtained new high signal-to-noise ratio spectrophotometric
observations for 12 BCDs with the
6.5 m MMT on the nights of 2010 April 13 and 2011
March 4 -- 5. In constructing the observational sample, we have selected 
from the SDSS DR7  
high-excitation H~{\sc ii} regions with strong
He~{\sc ii} $\lambda$4686 emission, i.e. those with   
$I$(He~{\sc ii} $\lambda$4686)/$I$(H$\beta$) $\ga$ 1--2\%. 
This selection criterion is motivated by the fact that  
the [Ne~{\sc v}] emission in Tol 1214$-$277, SBS 0335$-$052E and 
HS 0837$+$4717 is associated with relatively strong He~{\sc ii} emission. 
The general characteristics of 12 BCDs are listed in Table \ref{tab1} in order 
of increasing right ascension. 
All observations were made with the Blue Channel of the
MMT spectrograph. We used a 1\farcs5$\times$180\arcsec\ slit and a 800 grooves
mm$^{-1}$ grating in first order. The above instrumental setup gave
a spatial scale along the slit of 0\farcs6 pixel$^{-1}$, a scale perpendicular
to the slit of 0.75 \AA\ pixel$^{-1}$, a spectral range of 3200 -- 5200\AA\ 
and a spectral resolution of 3 \AA\ (FWHM). The seeing was $\sim$ 1\arcsec\ 
during both runs. 
The total exposure times varied between 45 and
60 minutes. Each exposure was broken up into 3 -- 4 subexposures,
not exceeding 15 minutes each, to allow for removal of cosmic
rays. Three Kitt Peak IRS spectroscopic standard stars, G191B2B,
Feige 34, and HZ 44 were observed at the beginning, middle,
and end of each night for flux calibration. Spectra of He+Ar
comparison arcs were obtained before or after each observation
to calibrate the wavelength scale.

The two-dimensional spectra were bias-subtracted and flat-field
corrected using IRAF.\footnote{IRAF is distributed by National Optical 
Astronomical Observatory, which is operated by the Association of Universities 
for Research in Astronomy, Inc., under cooperative agreement with the 
National Science Foundation.} We then use the IRAF software routines
IDENTIFY, REIDENTIFY, FITCOORD, and TRANSFORM
to perform wavelength calibration and correct for distortion and
tilt for each frame. Night sky subtraction was performed using
the routine BACKGROUND. The level of night sky emission
was determined from the closest regions to the galaxy that are free
of galaxian stellar and nebular line emission, as well as of emission
from foreground and background sources. Then, the two-dimensional spectra
were flux-calibrated using observations of standard stars. One-dimensional
spectra were finally extracted from each two-dimensional frame using
the APALL routine. 
The emission-line fluxes were measured
using the IRAF SPLOT routine. The line flux errors listed
include statistical errors derived with SPLOT from non-flux calibrated
spectra, in addition to errors introduced in the standard
star absolute flux calibration. Since the differences between the
response curves derived for the three standard stars are not greater
than 1\%, we set the errors in flux calibration to 1\% of the line
fluxes. The line flux errors will be later propagated into the
calculation of abundance errors. The line fluxes were corrected
for both reddening \citep{W58} and underlying hydrogen
stellar absorption derived simultaneously by an iterative procedure
as described in \citet{I94}. The extinction coefficient $C$(H$\beta$) for the 
line flux correction for interstellar dust is derived from the observed
decrement of the Balmer hydrogen emission lines. 

The corrected line
fluxes 100$\times$$I$($\lambda$)/$I$(H$\beta$), extinction
coefficients $C$(H$\beta$), equivalent width EW(H$\beta$) of the H$\beta$ 
emission line and equivalent widths EW(abs) of the hydrogen
absorption stellar lines are given in Table \ref{tab2}, along
with the uncorrected H$\beta$ fluxes.

\begin{table*}
 \centering
 \begin{minipage}{170mm}
  \caption{Spectroscopic characteristics.}\label{tab4}
  \begin{tabular}{@{}lccrlccclcc@{}}
  \hline
Name        &Redshift&12+logO/H&EW(H$\beta$)&\multicolumn{3}{c}{Flux}&log $L$(H$\beta$)&\multicolumn{3}{c}{FWHM$^{\rm c}$} \\ \cline{5-7} \cline{9-11}
            &        &         & \multicolumn{1}{c}{\AA}&H$\beta$$^{\rm a}$&4686$^{\rm b}$&3426$^{\rm b}$&&H$\beta$&4686&3426 \\
 \hline
\multicolumn{11}{c}{a) This paper} \\
J0905$+$0335  &0.03914 &7.77$\pm$0.01& 116.4& 0.76$\pm$0.01& 1.92$\pm$0.22& $<$0.55      &40.35&210&235&... \\
J0920$+$5234  &0.00780 &7.91$\pm$0.01& 139.1& 2.01$\pm$0.03& 0.97$\pm$0.10& $<$0.25      &39.37&176&192&... \\
J1016$+$3754  &0.00391 &7.57$\pm$0.01&  96.7& 4.74$\pm$0.08& 2.10$\pm$0.09& 0.58$\pm$0.15&39.14&188&207&228 \\
J1044$+$0353  &0.01287 &7.43$\pm$0.01& 257.8& 3.42$\pm$0.05& 2.01$\pm$0.08& 0.61$\pm$0.15&40.03&241&239&220 \\
J1050$+$1538  &0.08454 &7.94$\pm$0.01& 198.5& 2.28$\pm$0.04$^{\rm d}$& 1.60$\pm$0.32&$<$0.23&41.49&247$^{\rm e}$&325&... \\
J1053$+$5016  &0.00449 &7.98$\pm$0.01& 127.8& 2.55$\pm$0.04& 1.79$\pm$0.10& $<$0.17      &38.99&170&182&... \\
J1230$+$1202  &0.00408 &7.70$\pm$0.01&  84.5& 1.07$\pm$0.02& 3.81$\pm$0.20& 0.79$\pm$0.36&38.53&180&180&179 \\
J1323$-$0132  &0.02254 &7.61$\pm$0.02& 264.7& 0.63$\pm$0.01& 1.37$\pm$0.33& $<$0.39      &39.79&242&258&... \\
J1423$+$2257  &0.03287 &7.68$\pm$0.01& 129.0& 3.35$\pm$0.05& 3.23$\pm$0.15& 0.91$\pm$0.31&40.84&201&227&280 \\
J1426$+$3822  &0.02237 &7.78$\pm$0.01& 106.8& 0.80$\pm$0.01& 2.44$\pm$0.19& $<$0.31      &39.88&198&202&... \\
J1448$-$0110  &0.02733 &8.02$\pm$0.01& 134.7& 4.63$\pm$0.07& 0.86$\pm$0.13& $<$0.20      &40.82&258&264&... \\
J1545$+$0858  &0.03758 &7.70$\pm$0.01& 211.8& 3.03$\pm$0.04& 2.04$\pm$0.08& 0.51$\pm$0.16&40.91&198&228&190 \\ \hline
\multicolumn{11}{c}{b) Other galaxies with [Ne~{\sc v}] $\lambda$3426 emission} \\
SBS0335$-$052E$^{\rm f}$&0.01349 &7.31$\pm$0.01& 189.7& 10.2$\pm$0.15& 2.52$\pm$0.04& 0.72$\pm$0.06&40.55&227&260&371 \\ 
HS0837$+$4717$^{\rm f}$ &0.04195 &7.60$\pm$0.01& 235.0& 5.26$\pm$0.08& 1.99$\pm$0.07& 0.47$\pm$0.13&41.25&225&280&246 \\
Tol1214$-$277$^{\rm g}$ &0.02600 &7.56$\pm$0.01& 320.4& 3.13$\pm$0.05& 5.00$\pm$0.10& 2.70$\pm$0.40&40.61&...&...&... \\
\hline
\end{tabular}

$^{\rm a}$ Extinction-corrected flux in units 10$^{-14}$ erg s$^{-1}$ cm$^{-2}$.

$^{\rm b}$ Extinction-corrected fluxes in units 100$\times$$I$(4686)/$I$(H$\beta$) and 
100$\times$$I$(3426)/$I$(H$\beta$). 

$^{\rm c}$In km s$^{-1}$.

$^{\rm d}$From the SDSS spectrum.

$^{\rm e}$Width of the H$\gamma$ emission line.


$^{\rm f}$Data are from \citet{TI05}.

$^{\rm g}$Data are from \citet{I04a}.

\end{minipage}
\end{table*}

\section[]{Physical characteristics and element abundances}\label{sec:abund}

To determine element abundances, we follow generally the
procedures of \citet{I94} and \citet{I06a}. We adopt a two-zone 
photoionised H~{\sc ii} region model: a high-ionisation zone with temperature 
$T_{\rm e}$(O~{\sc iii}), where [O~{\sc iii}],
[Ne~{\sc iii}], and [Ar~{\sc iv}] lines originate, and a low-ionisation zone
with temperature $T_{\rm e}$(O~{\sc ii}), where [N~{\sc ii}], [O~{\sc ii}], 
[S~{\sc ii}], and [Fe~{\sc iii}] lines originate. 
As for the [S~{\sc iii}] and [Ar {\sc iii}] lines, they originate
in the intermediate zone between the high- and low-ionisation
regions.
The temperature $T_{\rm e}$(O~{\sc iii}) is calculated using the 
[O~{\sc iii}] $\lambda$4363/($\lambda$4959+$\lambda$5007) ratio. To take into 
account the electron temperatures for different ions, we have used the 
expressions of \citet{I06a}. Since our observations cover only the blue part
of the optical spectrum, the [S~{\sc ii}] $\lambda$$\lambda$6717, 6731 
emission lines usually used to determine the electron number density 
$N_{\rm e}$(S~{\sc ii}) were not available. Therefore, we set 
$N_{\rm e}$(S~{\sc ii}) = 10 cm$^{-3}$. The
low-density limit for abundance determinations should hold as
long as $N_{\rm e}$ is less than 10$^4$ cm$^{-3}$. Ionic and total heavy element
abundances for the 12 BCDs observed with the MMT are
derived in the manner described in \citet{I06a}. 
The electron temperatures $T_{\rm e}$(O~{\sc iii}), $T_{\rm e}$(O~{\sc ii}), ionic and
total heavy element abundances, the ionisation correction factors $ICF$ for
unseen stages of ionisation are shown in Table \ref{tab3}. 
The oxygen abundances are given in Table \ref{tab4}. They are 
generally low, in the range 
12 + logO/H = 7.43 -- 8.02, corresponding to heavy element mass fractions 
between 1/19 and 1/5 $Z_\odot$ with the solar calibration of \citet{A09}.
 The Ne/O and Fe/O abundance ratios are in the
range of abundance ratios for BCDs \citep[e.g. ][]{I06a}.

\begin{figure*}
\hbox{
\includegraphics[width=4.0cm,angle=-90.]{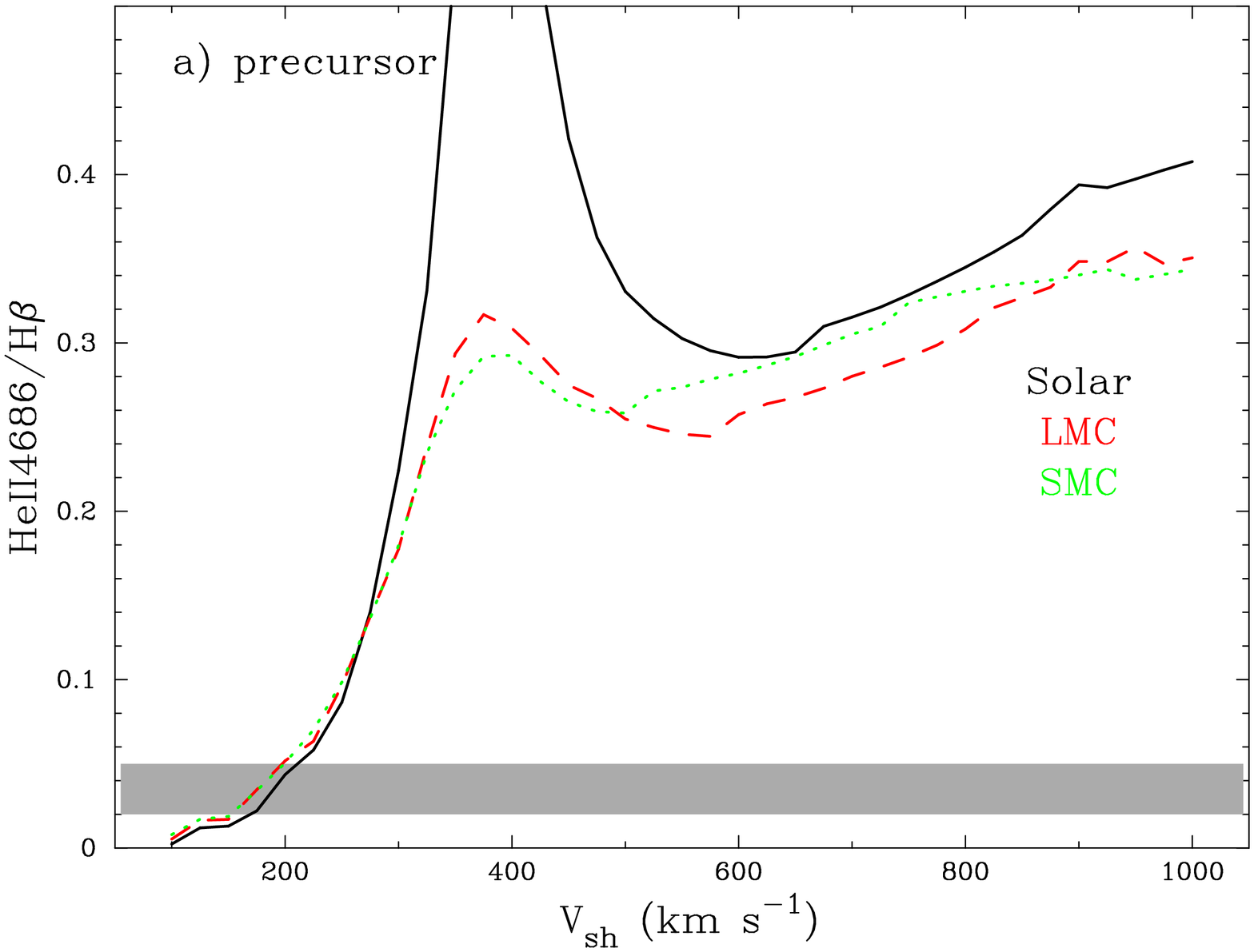}
\hspace{0.1cm}\includegraphics[width=4.0cm,angle=-90.]{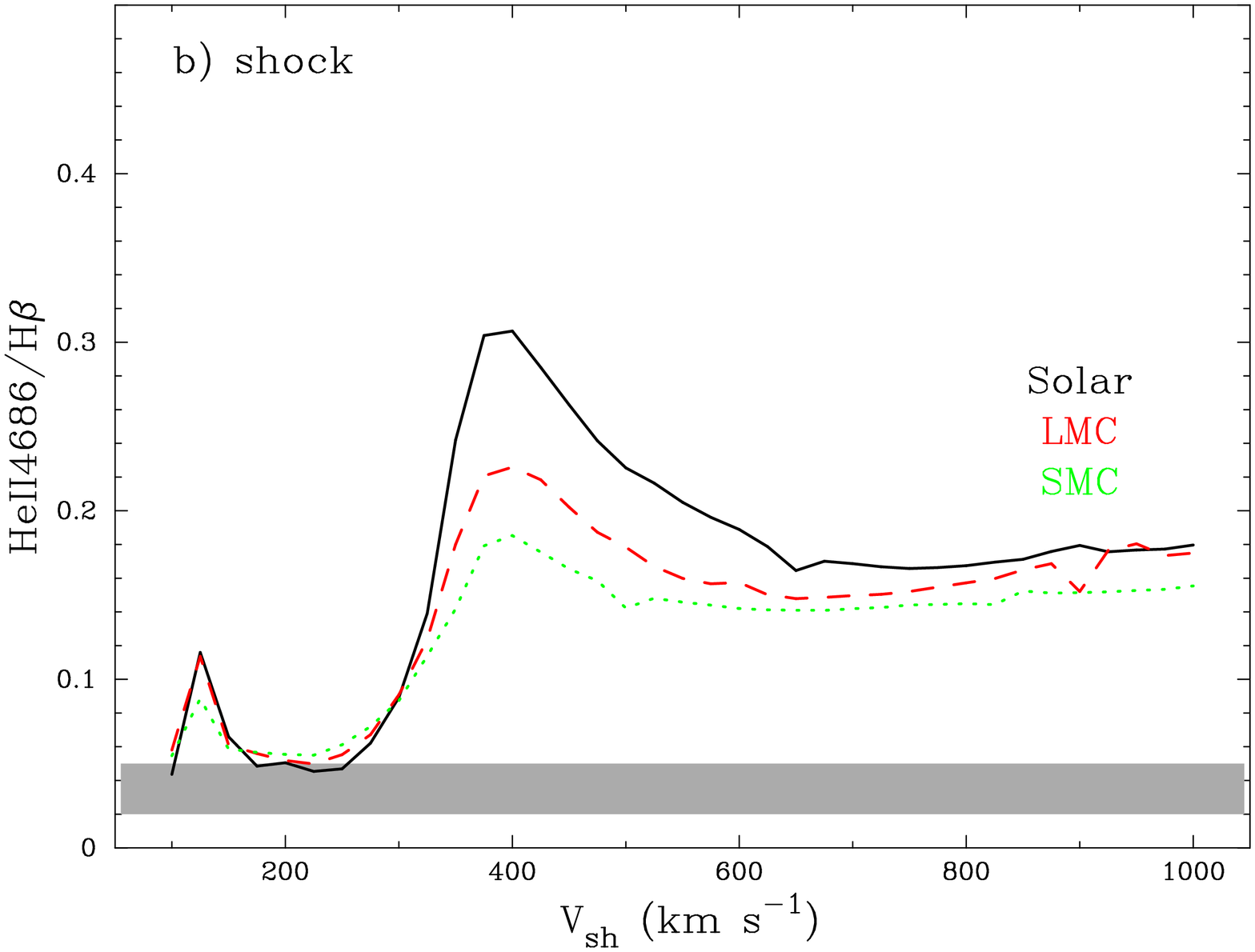}
\hspace{0.1cm}\includegraphics[width=4.0cm,angle=-90.]{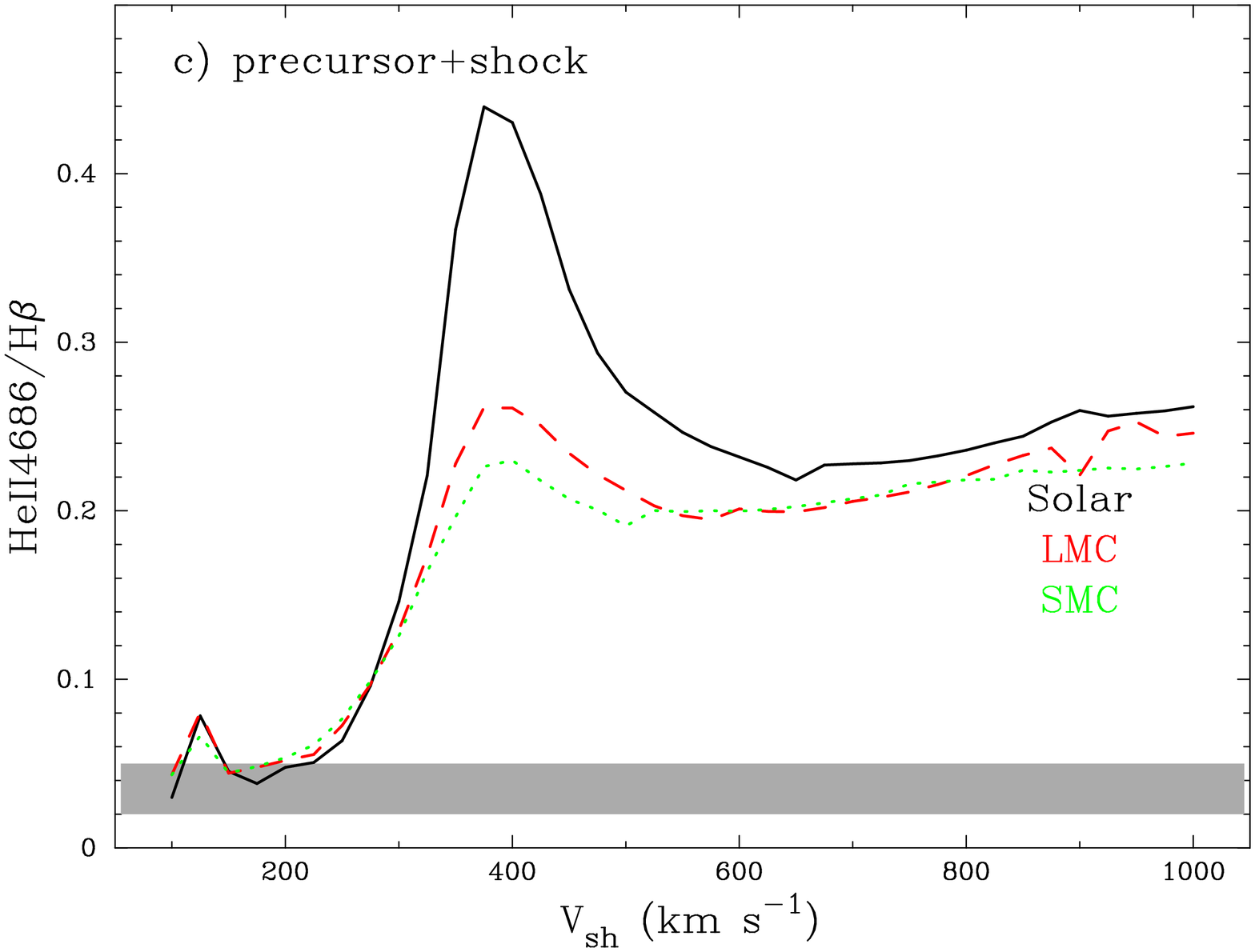}
}
\vspace{0.15cm}
\hbox{
\hspace{0.0cm}\includegraphics[width=4.0cm,angle=-90.]{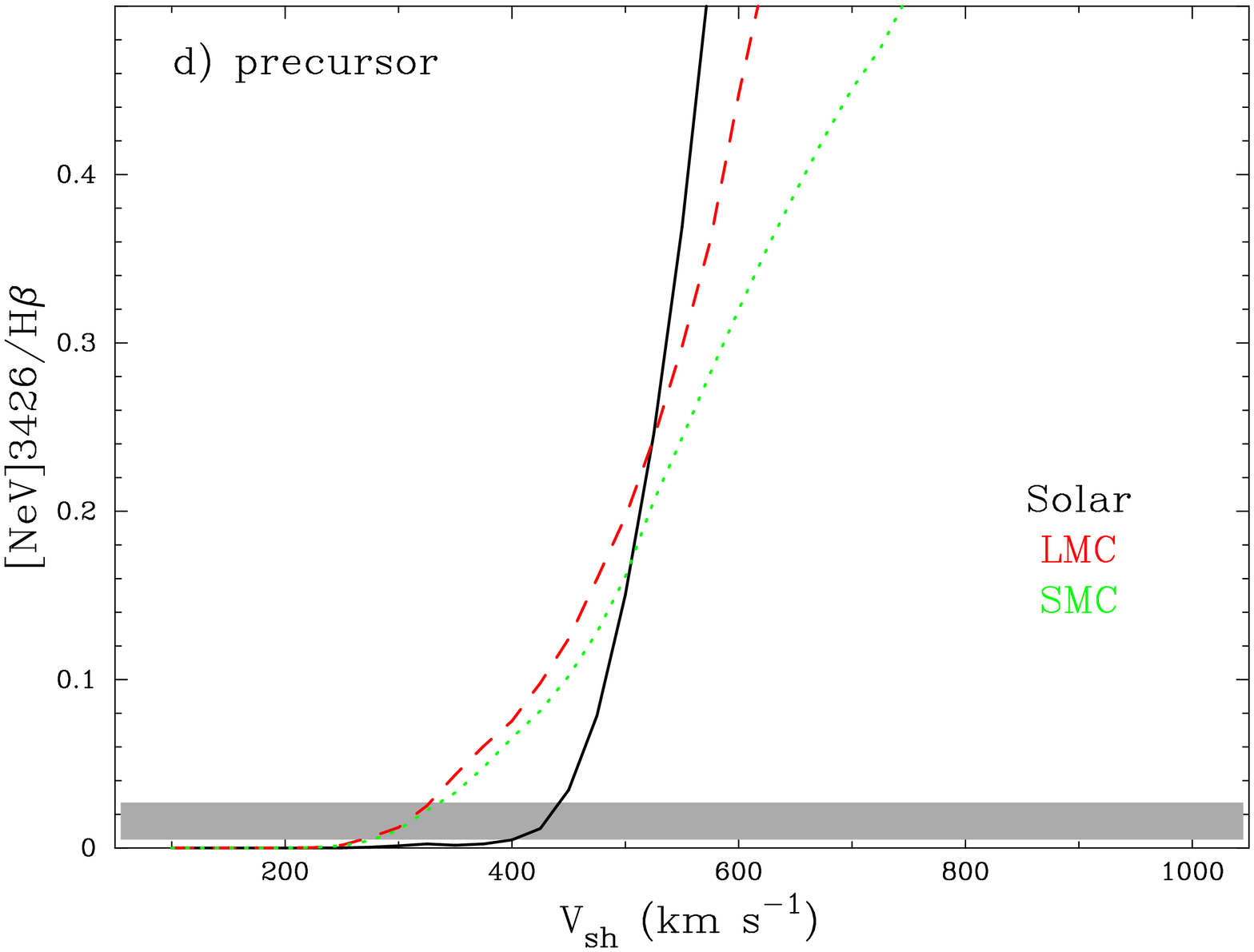}
\hspace{0.1cm}\includegraphics[width=4.0cm,angle=-90.]{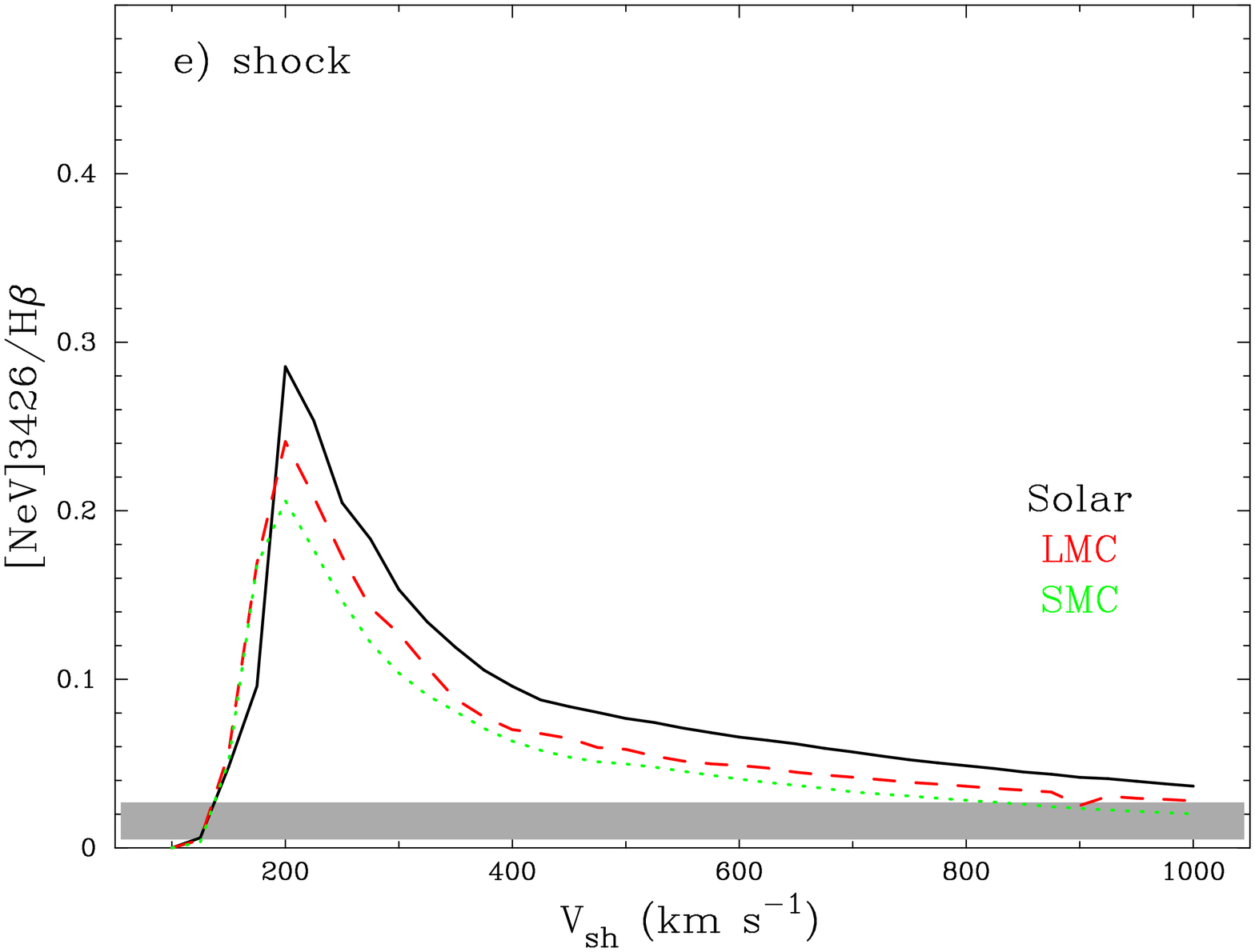}
\hspace{0.1cm}\includegraphics[width=4.0cm,angle=-90.]{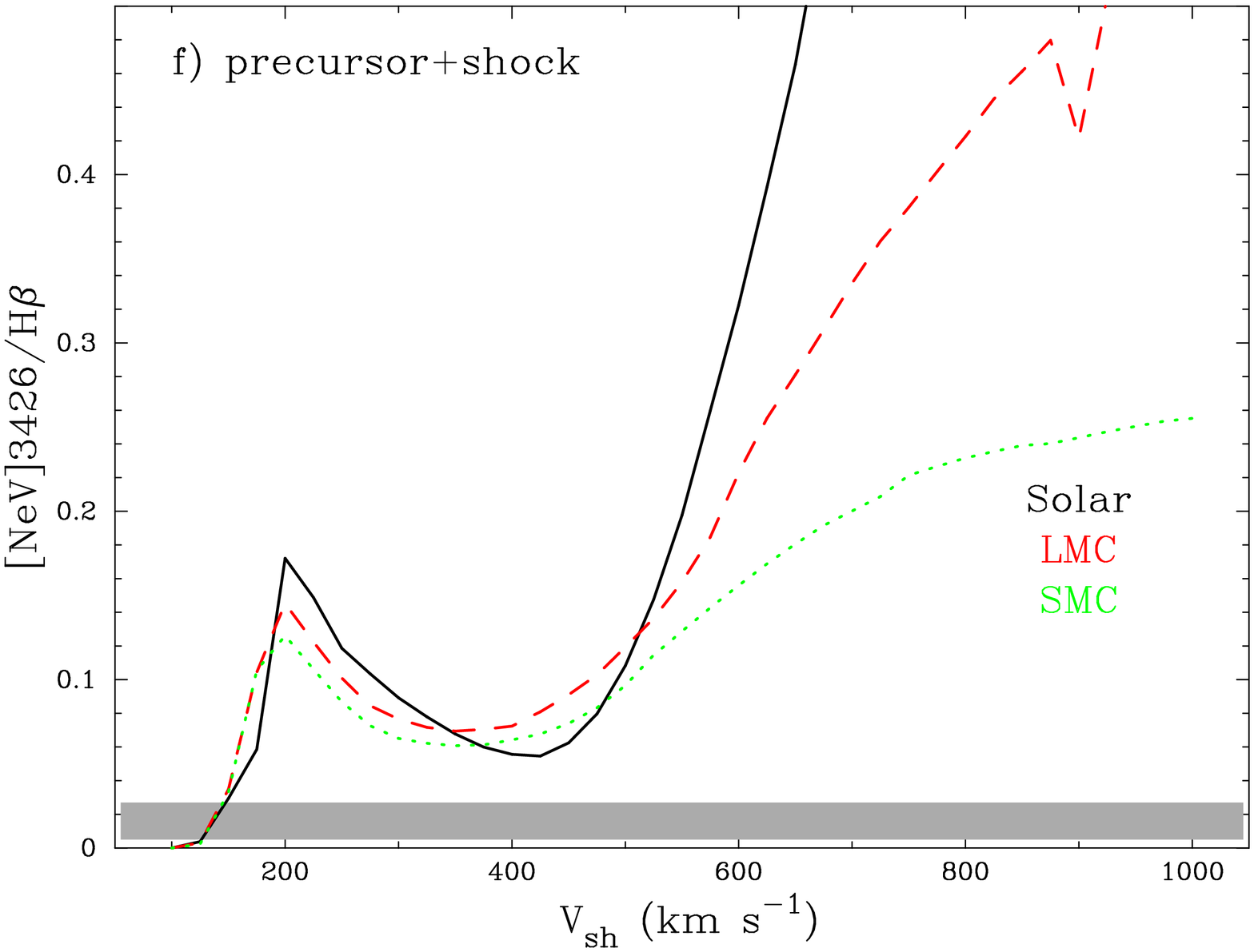}
}
\vspace{0.15cm}
\hbox{
\hspace{0.0cm}\includegraphics[width=4.04cm,angle=-90.]{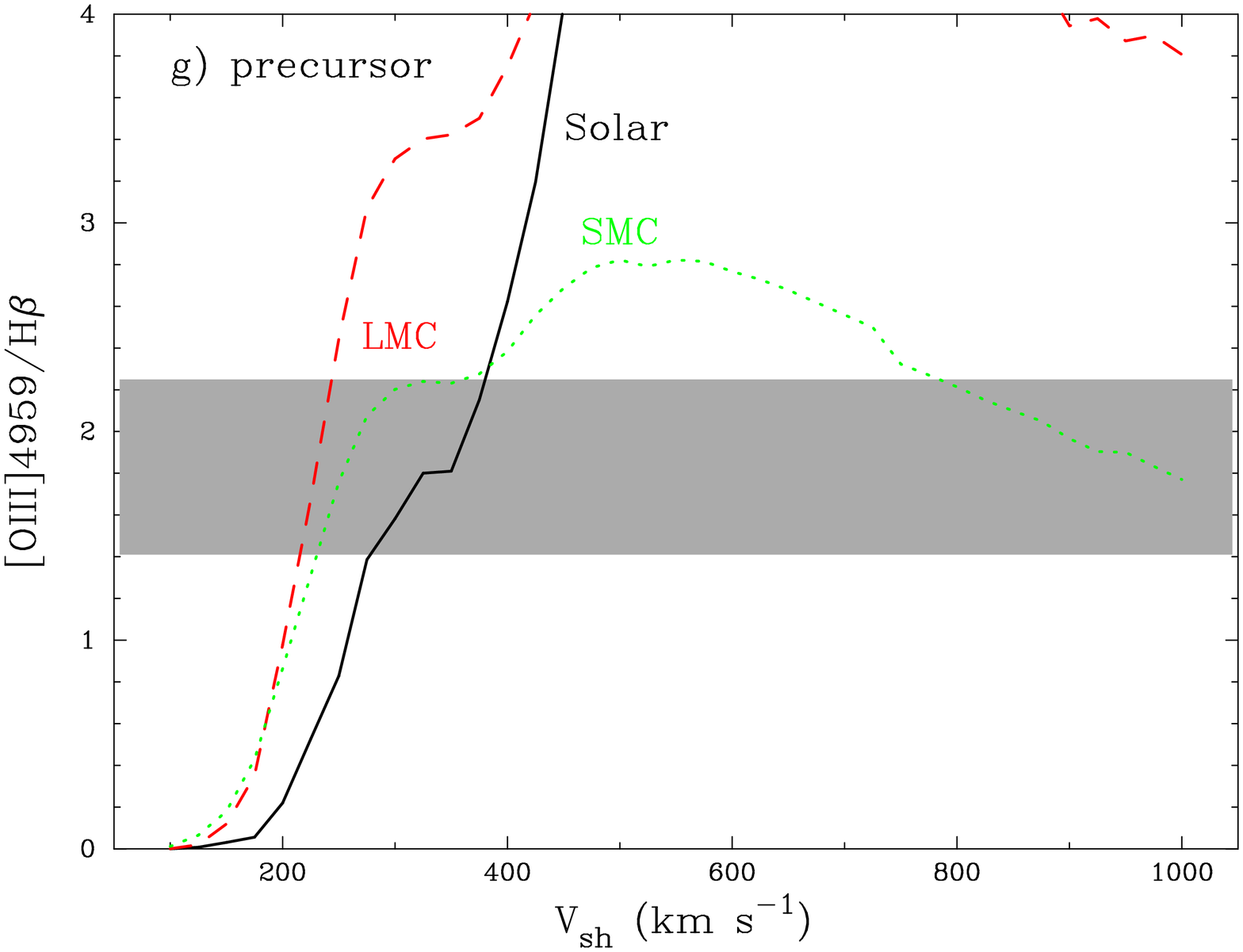}
\hspace{0.1cm}\includegraphics[width=4.04cm,angle=-90.]{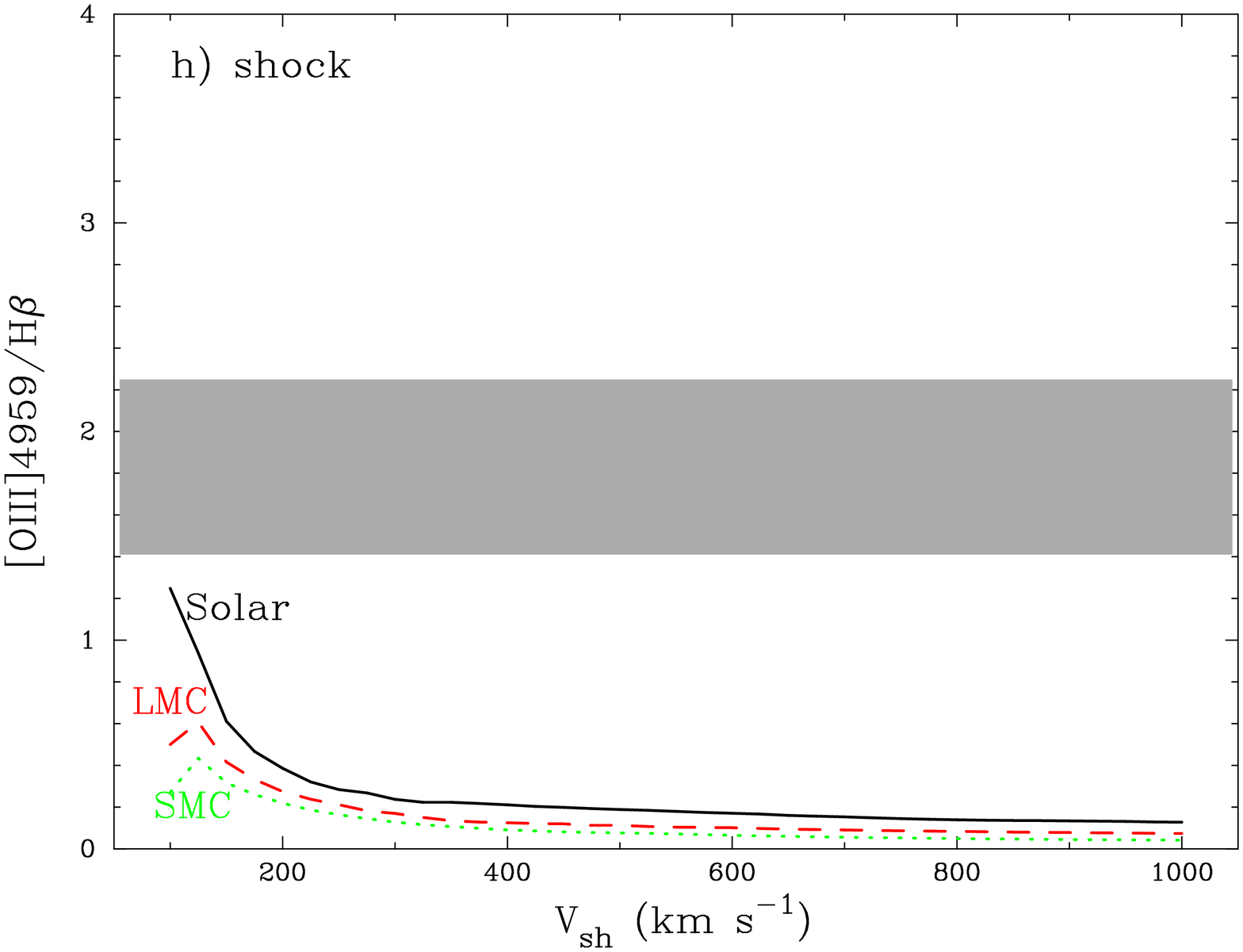}
\hspace{0.1cm}\includegraphics[width=4.04cm,angle=-90.]{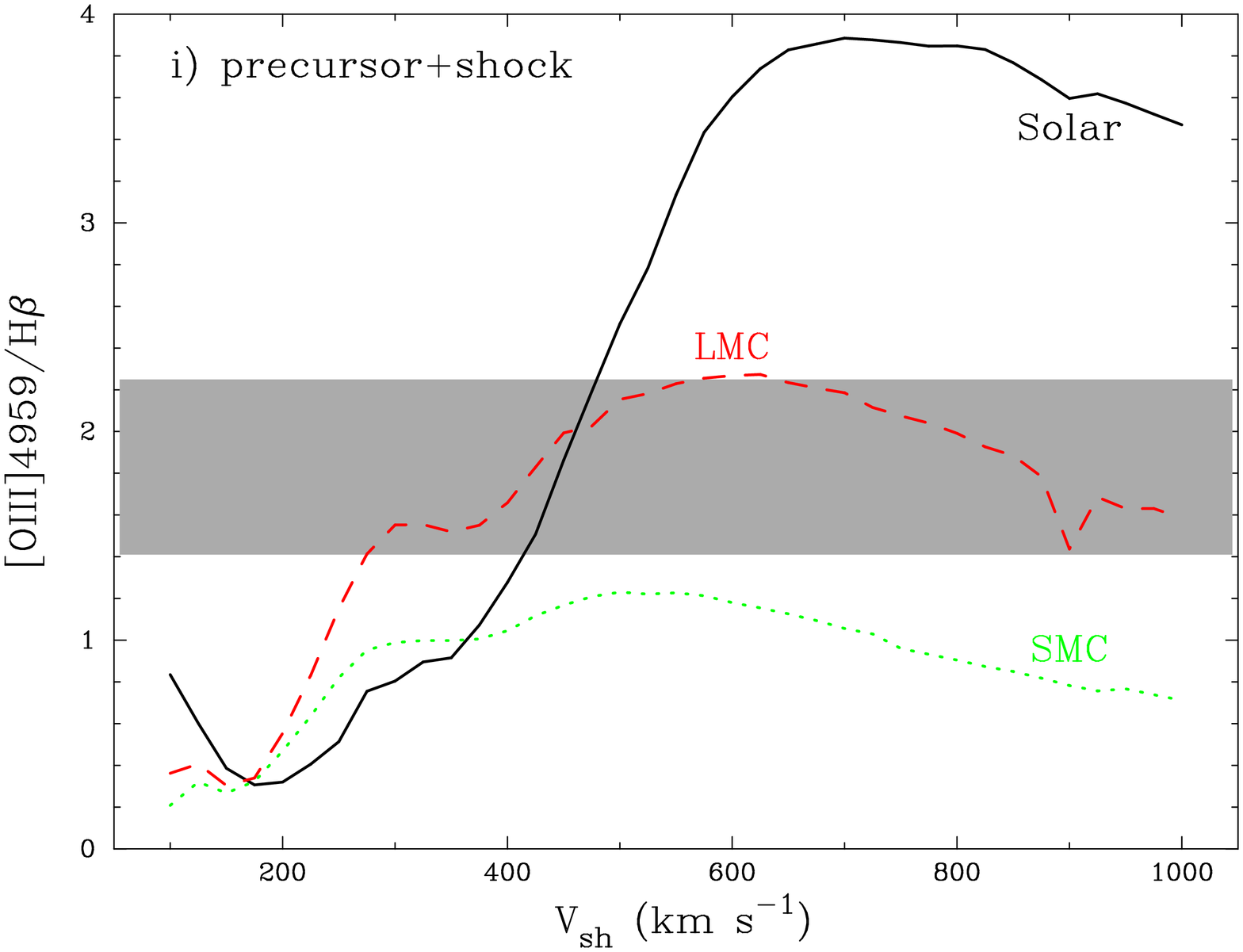}
}
\vspace{0.15cm}
\hbox{
\hspace{0.0cm}\includegraphics[width=4.08cm,angle=-90.]{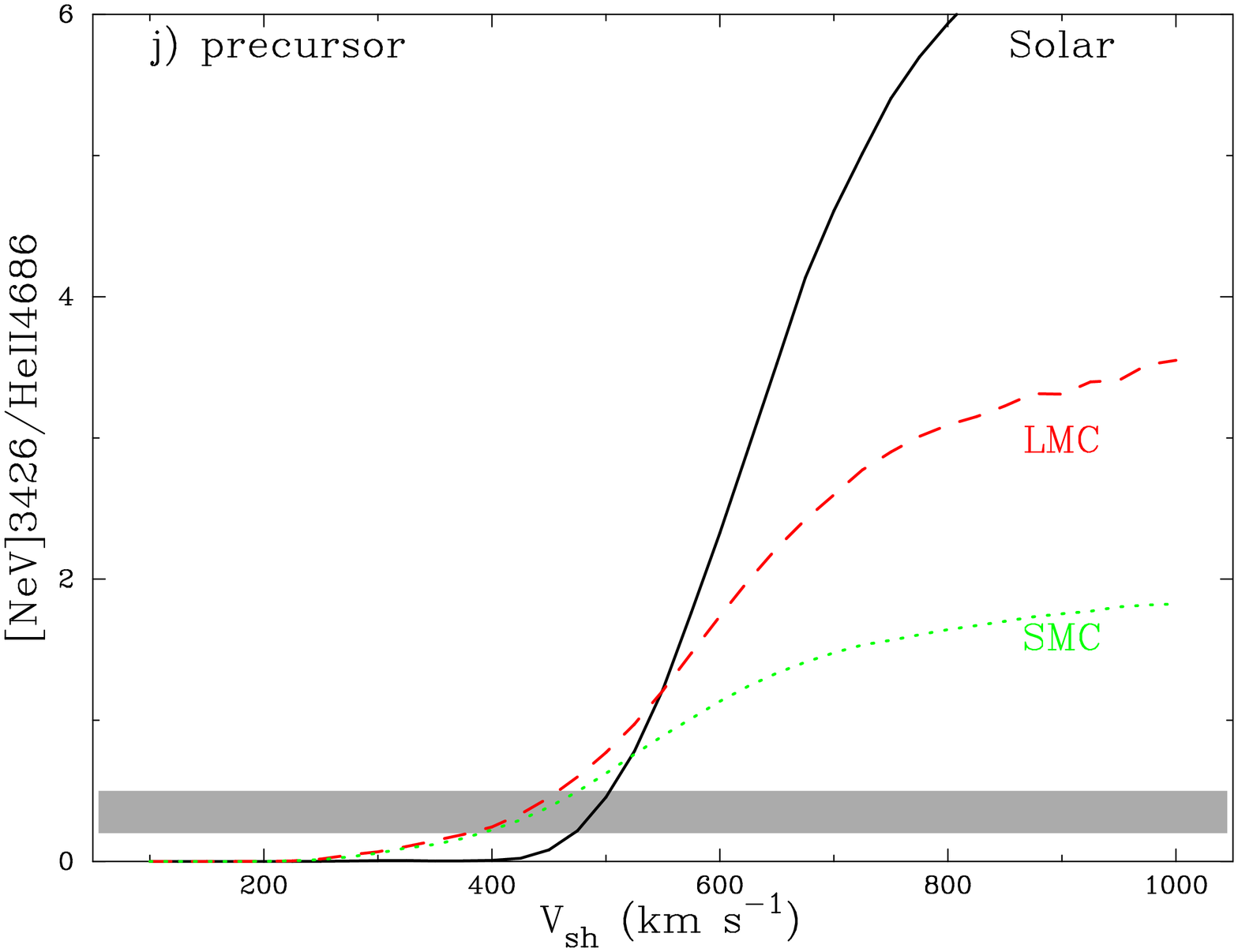}
\hspace{0.1cm}\includegraphics[width=4.08cm,angle=-90.]{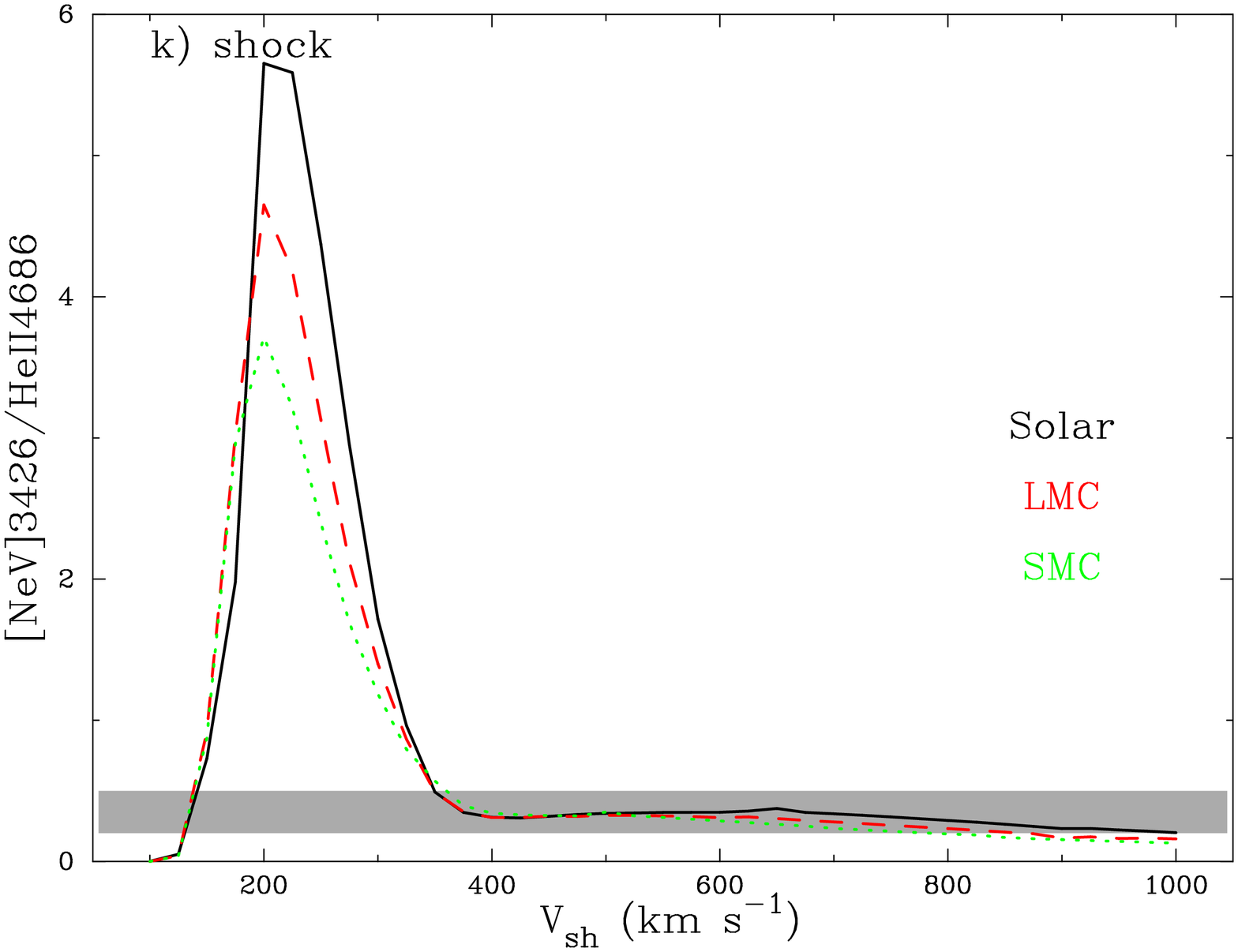}
\hspace{0.1cm}\includegraphics[width=4.08cm,angle=-90.]{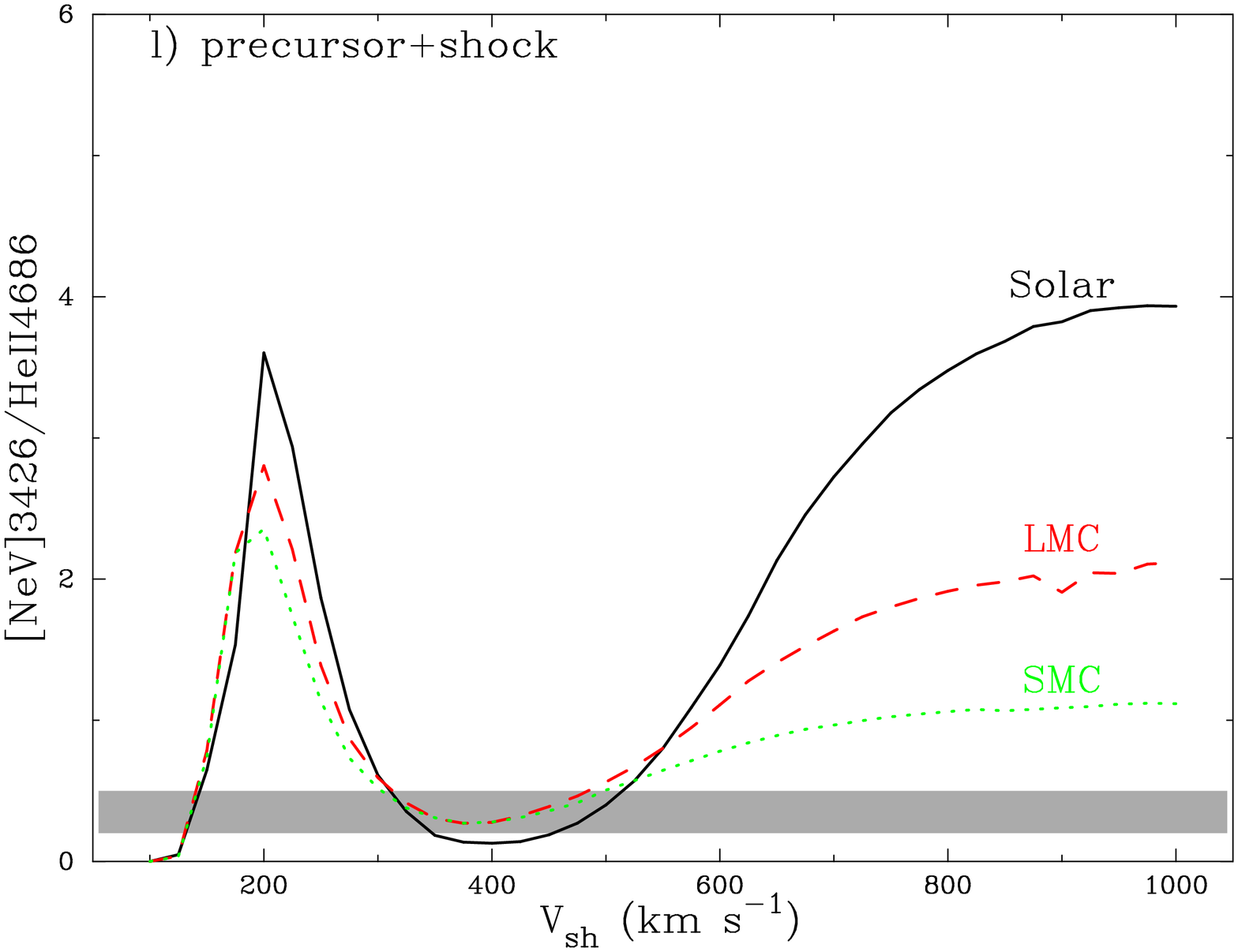}
}
 \caption{Dependences of various emission-line fluxes on the shock velocity 
$v_{\rm sh}$ according to the radiative shock models of \citet{A08}. Only data for  
shocks propagating through a medium with a number density 
$N_{\rm e}$ = 1 cm$^{-3}$ and a metallicity equal to that of the Sun (black solid lines), 
of the LMC (red dashed lines) and of the 
SMC (green dotted lines) are shown. From left 
to right are shown model predictions for the precursor, the shock and the
combined precursor+shock, respectively. The panels from up to bottom show  
the line flux ratios $I$(He~{\sc ii} 4686)/$I$(H$\beta$),  
$I$([Ne~{\sc v}] 3426)/$I$(H$\beta$), 
$I$([O~{\sc iii}] 4959)/$I$(H$\beta$), 
and $I$([Ne~{\sc v}] 3426)/$I$(He~{\sc ii} 4686), respectively.}
\label{fig3}
\end{figure*}

\section[]{Hard ionising radiation and origin of the [Ne~{\sc v}] $\lambda$3426
emission}\label{sec:hard}

\subsection{Observed properties}

The spectroscopic characteristics of the 12 observed BCDs are shown in 
Table \ref{tab4}.
It contains the galaxy name, the redshift $z$, the oxygen abundance 
12 + log O/H, the equivalent width EW(H$\beta$) of the H$\beta$ emission
line, the extinction-corrected fluxes of the H$\beta$, 
He~{\sc ii} $\lambda$4686 and 
[Ne~{\sc v}] $\lambda$3426 emission lines, the logarithm of the 
extinction-corrected H$\beta$ luminosity, and the full widths at half maximum
(FWHM) of the H$\beta$, He~{\sc ii} $\lambda$4686 and 
[Ne~{\sc v}] $\lambda$3426 emission lines. For comparison, we also
show data from the literature for the three BCDs known previously  
to have 
[Ne~{\sc v}] $\lambda$3426 emission. For the present sample,  
Table \ref{tab4} shows that [Ne~{\sc v}] $\lambda$3426 emission  
is present in five out of 12 BCDs, increasing the total number of known 
BCDs with [Ne~{\sc v}] $\lambda$3426 emission to eight. 
The SDSS images of the five BCDs with detected 
[Ne~{\sc v}] $\lambda$3426 emission in the present sample 
are shown in Fig. \ref{fig1}. All
of them have a compact structure with a bright H~{\sc ii} region. We note
however that all other observed BCDs without detected 
[Ne~{\sc v}] $\lambda$3426 emission line also have similar morphology, so 
morphology is not the discriminant factor. All BCDs with 
[Ne~{\sc v}] $\lambda$3426 emission are also dwarf galaxies, as seen by the 
absolute magnitudes listed in Table \ref{tab1}.

We show in Figure \ref{fig2} the redshift-corrected spectra of the 
five BCDs with detected [Ne~{\sc v}] $\lambda$3426 emission. 
The three dotted vertical lines show
respectively the locations of the [Ne~{\sc v}] $\lambda$3426, [Fe~{\sc v}]
$\lambda$4227, and He~{\sc ii} $\lambda$4686 emission lines. 
The [Fe~{\sc v}] $\lambda$4227 emission line is clearly detected in the 
three BCDs
J1044$+$0353, J1423$+$2257, and J1545$+$0858. The insets in each
panel show the expanded part of the spectra with the [Ne~{\sc v}] $\lambda$3426
emission line.

Inspection of Table \ref{tab4} shows that
the [Ne~{\sc v}] $\lambda$3426 line is detected only in BCDs with a 
He~{\sc ii} $\lambda$4686/H$\beta$
flux ratio $\ga$ 2\%. The situation is the same 
 in the three previously known BCDs with  
[Ne~{\sc v}] $\lambda$3426 emission \citep{I04b,TI05}. This implies 
that [Ne~{\sc v}] $\lambda$3426 and 
He~{\sc ii} $\lambda$4686 emission are associated. 
The 
[Ne~{\sc v}] $\lambda$3426/He~{\sc ii} $\lambda$4686 flux ratio 
in all eight BCDs is
in the range $\sim$ 20 -- 50\%. There is however one exception 
in Table \ref{tab4}: the galaxy J1426$+$3822 has a 
He~{\sc ii} $\lambda$4686/H$\beta$ flux ratio 
$>$ 2\%, yet no [Ne~{\sc v}] $\lambda$3426 emission line was detected. We
suggest that the non-detection is simply due to the faintness of the galaxy 
(it is fainter than all five galaxies with 
[Ne~{\sc v}] $\lambda$3426 emission),
resulting in a spectrum with a lower signal-to-noise ratio.
The BCDs with [Ne~{\sc v}] $\lambda$3426 emission tend 
to have lower oxygen abundances, in the range 7.3 -- 7.7, implying
hotter H~{\sc ii} regions. This is in line with the conclusions 
of \citet{GIT00} and \citet{TI05} who found that the nebular
He~{\sc ii} line emission is stronger in BCDs with lower metallicity, which 
would enhance 
the probability of detecting [Ne~{\sc v}] $\lambda$3426 emission. On the other
hand, there is no clear difference between 
the equivalent widths EW(H$\beta$) which measure the age of the starburst, and
luminosities $L$(H$\beta$) of the H$\beta$ emission line of BCDs 
with [Ne~{\sc v}] $\lambda$3426 emission and of those without. 

   We also do not
find any difference in the electron number density $N_{\rm e}$(S~{\sc ii}) (not shown in Table \ref{tab4}) between the two sets of galaxies. These are derived
from the [S~{\sc ii}] $\lambda$6717/$\lambda$6731 flux ratio in the SDSS
spectra, as the MMT spectra do not cover the 
red wavelength range.

\subsection{Origin of the hard radiation}

\subsubsection{AGN}

\citet{I04b} and \citet{TI05} have analysed some possible sources of the
hard ionising radiation responsible for the He~{\sc ii} and 
especially the [Ne~{\sc v}] emission. Although in normal galaxies, 
[Ne~{\sc v}] emission is usually attributed to the presence of an active 
galactic nucleus (AGN), those authors  
rule out such a presence because other emission lines 
usually associated with AGN activity are not seen. 
We have run CLOUDY (version v10.00) models \citep{F98} with an 
ionising spectrum which includes both 
stellar \citep{L99} and AGN \citep{MF87} 
ionising radiation. We found that,
to account for the observed strengths of both the [Ne~{\sc v}] $\lambda$3426
and He~{\sc ii} $\lambda$4686 emission lines, the number fraction of the AGN
ionising photons should be $\la$ 10\% of the number of stellar ionising 
photons.
Thus, we cannot rule out the presence of a non-thermal source of ionisation 
in the studied galaxies. 

\subsubsection{High-mass X-ray binaries}

The 
energy of the photons that produce Ne$^{4+}$ ions are in the 
extreme UV and soft X-ray range. 
Thus high-mass X-ray binaries (HMXBs) may play a role. \citet{I04b} have estimated that the X-ray luminosity required to reproduce the  [Ne~{\sc v}] emission in the BCD Tol1214--277 
at photon energies greater than 0.14 keV should be $L_X$ =10$^{39}$ -- 
10$^{40}$ ergs s$^{-1}$. Unfortunately, out of the 8 known galaxies with 
[Ne~{\sc v}] emission, only SBS 0335--052 has Chandra X-ray observations 
\citep{T04}. For this BCD, it is found that more than 90\% of its 0.5-10 keV 
flux comes from a point source with a luminosity of 3.5$\times$10$^{39}$ ergs 
s$^{-1}$. If that point source is composed of a single object, 
then its luminosity would place it in the range of the so-called 
ultraluminous X-ray sources (ULXs). However, \citet{TI05} did not 
consider HMXBs as the main mechanism for producing  [Ne~{\sc v}] emission.
The reason is that there are  
other BCDs, such as I Zw 18 \citep{T04} and Mrk 59 \citep{T12}
that are known to contain HMXBs, but do not show [Ne~{\sc v}] 
emission.    
 
\subsubsection{Stars}
    
Using photoionisation H~{\sc ii} models, \citet{SS99} have concluded that
both the He~{\sc ii} and [Ne~{\sc v}] emission in BCDs could be explained by 
the hard ionising radiation produced by WR stars. On the other hand,
\citet{TI05} have found that stellar radiation, 
including that of Wolf-Rayet stars, is too soft for producing 
He~{\sc ii} $\lambda$4686 emission with a flux above 0.1 -- 1\% that of 
H$\beta$. To resolve this disagreement, we have run a series of CLOUDY models
with pure stellar ionising radiation. We have adopted the ionising spectral
energy distribution from Starburst99 models with a heavy element mass fraction 
$Z$ = 0.001 \citep{L99}, corresponding to an oxygen abundance 12+logO/H
of $\sim$ 7.6. We find that a detectable He~{\sc ii} $\lambda$4686 emission
line is present only in H~{\sc ii} region models powered by starbursts
in the very short age range between 3.4 Myr and 3.6 Myr. Furthermore, the
maximum flux of this line is $<$ 0.5\% that of H$\beta$, several times
smaller than the observed flux. The discrepancy is considerably 
worse in the case of [Ne~{\sc v}] $\lambda$3426 emission. 
Moreover, the broad WR bump at $\lambda$4650 is present only in the
spectrum of J1423+2257. It is absent in the spectra of the other four BCDs 
with [Ne~{\sc v}] $\lambda$3426 emission. This is in line with
the conclusions of \citet{GIT00} and \citet{SB12} who also did not find 
a clear correlation between the presence of WR and nebular He {\sc ii}
emission.

\citet{TI05} found that it is possible, in principle, to reproduce
the observed He~{\sc ii} $\lambda$4686 and [Ne~{\sc v}] $\lambda$3426
emission line fluxes by models of very low metallicity ($Z$ $\la$
10$^{-7}$) massive ionising stars \citep{S02,S03}. 
However, such models of Population
III stars would predict equivalent widths of H$\beta$
emission line that are several times larger than those observed.
Thus, neither models of normal stars, of Wolf-Rayet stars, nor of primordial 
stars are able to reproduce the observed high-ionisation line fluxes.

\subsubsection{Fast radiative shocks}

One of the most promising explanations for the hard radiation and 
the high-ionisation emission lines in BCDs is the presence of the fast radiative
shocks produced by SNe. \citet{I04b} and \citet{TI05} using the
shock models of \citet{DS96} concluded that radiative shocks 
with velocities of $\sim$ 450 km s$^{-1}$ can account 
for the observed fluxes of both the He~{\sc ii} $\lambda$4686 and [Ne~{\sc v}]
$\lambda$3426 emission lines if the contribution of the shocks to the
observed flux of the H$\beta$ emission line is a few percent that of the
stars. However, only shocks propagating in an ISM with solar metallicity
were considered by \citet{DS96}.

Since the work of \citet{TI05}, a new grid of radiative shock models has been calculated by \citet{A08} for an ISM with solar, Large Magellanic Cloud (LMC) and Small Magellanic Cloud (SMC)  
metallicities, and a wide
range of shock velocities going from 100 to 1000 km s$^{-1}$. 
We now examine whether these models are able to reproduce 
the observed fluxes of the high-ionisation emission lines in our BCDs. 
However, a direct comparison of the models with the 
observations is not straightforward because:  
1) the dominant source of ionisation in BCDs is stellar emission, which 
would mask
the effect of shocks on the emission line fluxes; 
2) most of the shock models were calculated for shocks propagating in a 
neutral or weakly ionised ISM, while we expect shocks in BCDs to  
propagate in an ISM fully ionised by massive stellar clusters and
containing relatively high-ionisation species like O~{\sc iii}, Ne~{\sc iii},
Ar {\sc iv}. This is evidenced by the strong 
emission lines of these ions seen in the spectra
of the BCDs studied here (Fig. \ref{fig2}). We would then expect the
emission lines of low-ionisation species like [O~{\sc ii}] $\lambda$3727
not to be affected by shocks, as they may originate in regions 
different from those where shocks propagate. Therefore, these
cannot be used for shock diagnostics;
3) it is likely that the signatures of several shocks with 
different velocities are seen in the integrated spectra of BCDs.
Indeed, \citet{I96} and \citet{I07} have shown that the H$\beta$, [O~{\sc iii}] $\lambda$4959, $\lambda$5007, and H$\alpha$ 
emission lines in some BCDs exhibit broad components, with FWHMs of 
$\sim$ 1000 -- 2000 km s$^{-1}$ and fluxes 
of $\sim$ 1 -- 2\% of the narrow component fluxes. This broad emission 
is likely produced in the adiabatic shocks 
propagating with velocities of several thousand km s$^{-1}$ 
\citep[e. g. ][]{C77}. Similar broad components of H$\beta$, 
and [O~{\sc iii}] $\lambda$4959, $\lambda$5007 are observed in the spectra
of some of the BCDs studied in this paper, such as J1044+0353 and 
J1545+0858 (Fig. \ref{fig2}).
On the other hand, the He~{\sc ii} and [Ne~{\sc v}]
emission likely arises in the preshock and compressed postshock regions of
slower radiative shocks.

In Fig. \ref{fig3}, we show from top to bottom the predicted 
$I$(He~{\sc ii} $\lambda$4686)/$I$(H$\beta$),
$I$([Ne~{\sc v}] $\lambda$3426)/$I$(H$\beta$),
$I$([O~{\sc iii}] $\lambda$4959)/$I$(H$\beta$), and
$I$([Ne~{\sc v}] $\lambda$3426)/$I$(He~{\sc ii} $\lambda$4686)
emission line flux ratios
for the precursor, shock and combined precursor+shock
models (from left to the right) as a function of the shock velocity $v_{\rm sh}$.
Models with solar, LMC and SMC metallicities are shown by a black solid line,
a red dashed line, and a green dotted line, respectively. All models have 
a number density of the ambient gas $N_{\rm e}$ = 1 cm$^{-3}$, as the grid of models 
of \citet{A08} for different metallicities
was calculated only for that density. 
However, we have checked from the solar metallicity models, which 
were calculated for a range of $N_{\rm e}$,
that the dependence of the emission line flux ratios on $N_{\rm e}$ is weak in 
the range $N_{\rm e}$ = 0.1 -- 10$^3$ cm$^{-3}$. The shaded regions indicate 
the range of the observed emission line flux ratios in the BCDs with detected 
[Ne~{\sc v}] $\lambda$3426 emission.

Inspection of Fig. \ref{fig3} shows that the predicted 
$I$(He~{\sc ii} $\lambda$4686)/$I$(H$\beta$) and
$I$([Ne~{\sc v}] $\lambda$3426)/$I$(H$\beta$) line ratios
are generally 
considerably larger than the observed ones.  
Agreement is only possible in the very narrow range of
shock velocities, $v_{\rm sh}$ $<$ 200 km s$^{-1}$. 
To have agreement for higher shock velocities, one would have  
to assume that the
contribution of the shock ionising radiation to the excitation of the
H$\beta$ emission line is small, not exceeding $\sim$ 10\% that of the
stellar ionising radiation. This would be a natural assumption 
since the studied BCDs contain numerous hot O stars,  
producing copious 
amounts of ionising photons. 
There is another problem with the shock models with 
$v_{\rm sh}$ $<$ 200 km s$^{-1}$. While the modelled line flux ratios in 
Fig. \ref{fig2}c, \ref{fig2}f, and \ref{fig2}l are comparable to the observed
ratios, implying that shocks can be the only source of ionisation, there 
is no such good agreement between the  
predicted and observed emission line ratios of other species.
In particular, the predicted
[O~{\sc iii}] $\lambda$4959/H$\beta$ flux ratios (Fig. \ref{fig2}i) are 
several times lower than the observed ones. In fact, the shock models
with the SMC oxygen abundance of 12+logO/H = 8.1 underpredict this 
ratio over the entire range of shock
velocities. For the lower oxygen abundances that are characteristic of our BCDs,
the discrepancy would be even worse.
Therefore, we conclude that the dominant source
of ionisation for the majority of the species in the BCDs studied here 
is stellar radiation, with the exception
of the high-ionisation species which would require a $\sim$ 10\% contribution 
from shock radiation.

The observed 
$I$([Ne~{\sc v}] $\lambda$3426)/$I$(He~{\sc ii} $\lambda$4686) flux 
ratio is in the range 0.2 -- 0.5 (Table \ref{tab4}). If a single shock is
assumed, then this range corresponds to shock velocities of $\sim$ 300 -- 500
km s$^{-1}$ (Fig. \ref{fig3}l). However, the situation is more complicated 
when there are several shocks superposed on one another. This is 
to be expected in our BCDs which contain 
a large number of massive stars and hence many supernova remnants. The modelled 
He~{\sc ii} $\lambda$4686/H$\beta$ and
[Ne~{\sc v}] $\lambda$3426/H$\beta$ flux ratios would be both in
agreement with the observed ratios if the fraction of the H$\beta$
flux produced by shocks is $\sim$10\% that produced by stellar ionising 
radiation, in the case of LMC and SMC metallicities 
(Fig. \ref{fig3}c and \ref{fig3}f), and if 
shocks velocities are in the 300 -- 500 km s$^{-1}$ range. These shocks 
produce an appreciable
amount of extreme UV photons, but at higher X-ray energies of $\sim$ 1 kev,
the ionising flux is predicted to 
decrease considerably \citep[e. g., Fig. 2 in ][]{A08}. Therefore, no
appreciable X-ray emission is expected from our BCDs. Unfortunately, 
we cannot check this prediction as none of the 
BCDs in Table \ref{tab4} have X-ray observations, 
except for the
brightest one, SBS 0335-052E, where weak diffuse X-ray emission, 
accounting for 10\% of the total X-ray emission, was detected. The other 90\% 
come from an ultraluminous X-ray binary, as described before
\citep{T04}.

\subsubsection{Preshock and shock regions} 

As for the preshock (precursor) and shock
regions, they contribute comparably to the excitation of He~{\sc ii} and 
[Ne~{\sc v}] emission, for $v_{\rm sh}$ $\sim$ 300 - 500 km s$^{-1}$. However,
we would expect a width difference in the lines arising 
from these regions. The preshock
region is less disturbed, so the widths of the lines produced in that region
should be narrow, of $\la$ 50 -- 100 km s$^{-1}$. On the other hand,
lines produced in the shocked region would be broader. However, the widths
of these lines would depend on the characteristic time of thermalisation,
resulting in the highest width if the postshock region is thermalised.
In any case, the postshock thermal velocities would be lower than the shock 
velocity.

In Table \ref{tab4}, we show the FWHMs of the H$\beta$, He~{\sc ii} $\lambda$4686 
and [Ne~{\sc v}] $\lambda$3426 emission lines. 
They are all very similar,  
$\sim$ 200 -- 250 km s$^{-1}$, corresponding to a velocity dispersion 
$\sigma$ = FWHM/2.3 $\la$ 100 km s$^{-1}$, significantly lower than the
shock velocity. However, we point out that the emission lines in the MMT spectra
are nearly unresolved, so that the measured widths are mainly instrumental. 
\citet{I06b}, using higher resolution VLT/GIRAFFE observations
of SBS 0335$-$052E, have measured the FWHMs of 
H$\beta$ and He~{\sc ii} $\lambda$4686 to be respectively  
$\sim$ 90 km s$^{-1}$ and $\sim$ 140 km s$^{-1}$. This 
difference in the FWHM is natural, as the H$\beta$ emission line is produced 
mainly in the undisturbed gas ionised by stars, while 
He~{\sc ii} $\lambda$4686 emission arises both in 
the preshock and postshock ISM. Unfortunately,
the observations of \citet{I06b} do not cover the near-UV range, and hence no
high spectral resolution data for [Ne~{\sc v}] $\lambda$3426 are available.

\section[]{Conclusions}\label{sec:concl}

We present MMT spectroscopic observations in the wavelength range 
3200 - 5200\AA\ of 12 blue compact dwarf (BCD) galaxies.
Our aim
was to detect the high-ionisation emission line [Ne~{\sc v}] $\lambda$3426\AA\
and to study the excitation mechanisms of this emission. Our main results
are as follows:

1. The [Ne~{\sc v}] $\lambda$3426\AA\ emission line was detected in five 
BCDs. Adding the three BCDs known previously to have [Ne~{\sc v}] emission 
\citep{I04b} \citep{TI05}, there is  
now a total of 8 BCDs known to possess this property.

2. [Ne~{\sc v}] $\lambda$3426\AA\ emission was found only in 
galaxies with low oxygen abundance (12+logO/H = 7.3 -- 7.7) and showing
relatively strong He~{\sc ii} $\lambda$4686\AA\ emission ($\ga$ 2\%  
of the H$\beta$ emission line flux). The  
[Ne~{\sc v}] $\lambda$3426/He~{\sc ii} $\lambda$4686 flux ratio
ranges between 0.2 and 0.5.
On the other hand, there is no significant correlation of the [Ne~{\sc v}] 
flux with the equivalent width EW(H$\beta$) and the luminosity
$L$(H$\beta$) of the H$\beta$ emission line.

3. The [Ne~{\sc v}] and He~{\sc ii} emission in the BCDs cannot be produced by
ionising radiation from high-mass X-ray binaries or 
massive main sequence stars, in agreement with the conclusions
from previous studies. Wolf-Rayet stars are also ruled out. 
Out of five BCDs with detected 
[Ne~{\sc v}] $\lambda$3426\AA\ emission, broad WR emission was 
detected only in one galaxy.

4. The observed 
[Ne~{\sc v}] $\lambda$3426/He~{\sc ii} $\lambda$4686 flux ratio can be 
reproduced in models with either AGN or 
radiative shock ionising radiation.  In the case of AGN, the number 
fraction of the AGN ionising photons should be $\la$10\% of the number 
of stellar ionising photons. 
 
Another likely ionising source for the [Ne~{\sc v}] and He~{\sc ii}
emission is shocks produced by  supernovae. However, comparison with 
shock models is not
straightforward because: 1) we likely observe the superposition of an ensemble 
of SNe remnants with various shock velocities while shock models are 
calculated for one given shock velocity; and 
2) the main source of ionisation
of the lower ionisation species (including hydrogen) 
is stellar radiation, thus excluding 
a direct comparison of the observed flux ratios of both high- and 
low-ionisation 
species with shock model predictions. Therefore, the only constraint on 
shock models is the [Ne~{\sc v}] $\lambda$3426/He~{\sc ii} $\lambda$4686 flux 
ratio. We find that the observed range of this flux ratio is consistent with
models of radiative shocks if shock velocities are in the  
300 -- 500 km s$^{-1}$ range, and if the shock ionising contribution 
is also $\sim$10\% of the stellar ionising contribution.  
 This conclusion is in agreement with previous findings
by \citet{I04b} and \citet{TI05}. Faster shocks, likely adiabatic, with
velocities $\ga$ 1000 -2000 km s$^{-1}$ are also present in our 
BCDs, as suggested by the low-intensity broad components of the H$\beta$ 
and [O~{\sc iii}] $\lambda$4959, 5007\AA\ emission lines seen in some objects.

\section*{Acknowledgments}

Y.I.I. thanks Dr Grazyna Stasi\'nska for helpful suggestions. T.X.T. thanks 
the hospitality of the Institut d'Astrophysique de Paris.

\bsp

\label{lastpage}

\end{document}